\documentclass[aps,prd,onecolumn,showpacs,groupedaddress]{revtex4-2}
\usepackage{graphicx}
\usepackage{dcolumn}
\usepackage{bm}
\usepackage{color}
\usepackage{longtable}
\usepackage{supertabular}
\usepackage[T1]{fontenc}
\usepackage{epstopdf}
\usepackage{amsmath}
\usepackage{amssymb}
\usepackage{setspace}
\usepackage{multirow}
\usepackage{booktabs}
\usepackage[section]{placeins}
\usepackage{mathrsfs}
\usepackage{hyperref}

\makeatletter

\newcommand{\Rmnum}[1]{\expandafter\@slowromancap\romannumeral #1@} 
\newcommand{\bq}{\begin{equation}}
\newcommand{\eq}{\end{equation}}
\newcommand{\bqn}{\begin{eqnarray}}
\newcommand{\eqn}{\end{eqnarray}}
\newcommand{\nb}{\nonumber}
\newcommand{\lb}{\label}

\makeatother

\begin{document}

\title{An MLE-based analysis of flow factorization and event-plane correlations}

\author{Chong Ye$^{1,2}$}
\author{Cesar A. Bernardes$^{3}$}
\author{Wei-Liang Qian$^{4,2,1}$}\email[E-mail: ]{wlqian@usp.br (corresponding author)}
\author{Sandra S. Padula$^{5}$}
\author{Rui-Hong Yue$^{1}$}
\author{Yogiro Hama$^{6}$}
\author{Takeshi Kodama$^{7,8}$}

\affiliation{$^{1}$ Center for Gravitation and Cosmology, College of Physical Science and Technology, Yangzhou University, 225009, Yangzhou, China}
\affiliation{$^{2}$ Faculdade de Engenharia de Guaratinguet\'a, Universidade Estadual Paulista, 12516-410, Guaratinguet\'a, SP, Brazil}
\affiliation{$^{3}$ Instituto de F\'isica, Universidade Federal do Rio Grande do Sul, 91501-970, Porto Alegre, RS, Brazil}
\affiliation{$^{4}$ Escola de Engenharia de Lorena, Universidade de S\~ao Paulo, 12602-810, Lorena, SP, Brazil}
\affiliation{$^{5}$ Instituto de F\'isica Te\'orica, Universidade Estadual Paulista, 01140-070, S\~ao Paulo, SP, Brazi}
\affiliation{$^{6}$ Instituto de F\'isica, Universidade de S\~ao Paulo, 05315-970, S\~ao Paulo, SP, Brazil}
\affiliation{$^{7}$ Instituto de F\'isica, Universidade Federal do Rio de Janeiro, 21945-970, Rio de Janeiro-RJ , Brazil}
\affiliation{$^{8}$ Instituto de F\'isica, Universidade Federal Fluminense, 24210-346, Niter\'oi, RJ, Brazil}

\begin{abstract}
In this study, we use the maximum likelihood estimator (MLE) to explore factorization and event-plane correlations in relativistic heavy-ion collisions.
Our analyses incorporate both numerical simulations and publicly available data from the CMS Collaboration.
We focus on Au+Au collisions at 200 GeV and Pb+Pb collisions at 2.76 TeV.
The differential flows obtained for various centrality windows and momentum cuts are consistent with conventional methodologies, such as multi-particle cumulants and event-plane methods.
Leveraging these findings, we proceed to undertake further analysis of flow factorization and event-plane correlations.
These quantities are relevant because of their sensitivity to initial-state fluctuations.
While higher-order correlators might provide different implementations of factorization ratio, the MLE is readily applied to these scenarios.
Moreover, the MLE's unique capabilities allow us to compute specific correlators that are typically inaccessible by other means.
As an asymptotically normal and unbiased estimator, the MLE provides a valuable alternative tool for flow and correlation analyses.
\end{abstract}

\date{Jan. 31th, 2025}

\maketitle
\newpage

\section{Introduction}\label{section1}

Relativistic hydrodynamics stands out as one of the viable theoretical frameworks for describing the temporal evolution of the strongly coupled quark-gluon plasma (QGP) generated in relativistic heavy-ion collisions~\cite{hydro-review-04, hydro-review-05, hydro-review-06, hydro-review-07, hydro-review-08, hydro-review-09, hydro-review-10}.
This macroscopic approach models the QGP as a continuum, essential for understanding the underlying physics that leads to observable phenomena.
The primary observables pertinent to the relativistic hydrodynamics are the particle spectrum observed at intermediate and low transverse momentum, and especially those related to the collective properties such as flow harmonics and correlations~\cite{Ollitrault:1992bk, Voloshin:1994mz, Ollitrault:1997vz, Borghini:2000sa, Takahashi:2009na, Andrade:2010xy, Luzum:2010sp}.
From an experimental standpoint, the anisotropic azimuthal distributions measured have been pivotal in establishing the concept of a {\it perfect} liquid, first evidenced at RHIC~\cite{RHIC-brahms-overview-1, RHIC-phobos-overview-1, RHIC-star-overview-1, RHIC-phenix-overview-1}.
Consequently, analyzing the azimuthal anisotropy of nuclear collisions has become a key observable to extract critical information about the properties of the underlying physical system~\cite{BRAHMS:2004adc, PHOBOS:2004zne, STAR:2005gfr, ATLAS:2012at, CMS:2012zex, CMS:2012tqw}.
Recently, these quantities have been further scrutinized for deformed nuclei~\cite{Holtermann:2023vwr, ATLAS:2014qxy, ATLAS:2019peb}.

The anisotropic azimuthal distribution of particles in momentum space is characterized by the flow harmonics $v_n$.
Specifically, the one-particle distribution function is given by~\cite{Voloshin:1994mz, Voloshin:2008dg}
\begin{eqnarray}
f_1(\phi)=\frac{1}{2\pi}\left[1+\sum_{n=1}2v_{n}\cos{n(\phi-\Psi_n)}\right] ,
\label{oneParDis}
\end{eqnarray}
where $\phi$ is the azimuthal angle of an emitted particle, and $\Psi_n$ denotes the event plane of order $n$.
Elliptic flow $v_2$ and triangular flow $v_3$ are particularly relevant, with $v_2$ primarily arising from the initial system's geometric almond shape of the overlapping nuclei~\cite{Ollitrault:1992bk}, and $v_3$ resulting from event-by-event fluctuations in the initial conditions~\cite{Alver:2010gr}.
Considerable research has been dedicated to exploring the relationship between initial geometric anisotropy and final-state flow harmonics, mainly focusing on deviations from linearity~\cite{Teaney:2012ke, Niemi:2012aj, Qian:2013nba, Yan:2014nsa, Fu:2015wba, Yan:2015jma, Wen:2020amn}, eccentricity and flow fluctuations~\cite{Hama:2007dq, Bhalerao:2011yg, Heinz:2013bua, Gronqvist:2016hym}, and correlations~\cite{Bhalerao:2013ina, Denicol:2014ywa}.

Various methods have been developed to determine the flow harmonics $v_n$ from experimental data numerically.
The traditional event plane method~\cite{Voloshin:1994mz, Poskanzer:1998yz} estimates the event planes $\Psi_n$ in Eq.~\eqref{oneParDis} to evaluate the flow harmonics.
This method is closely related to the fact that the reaction plane~\cite{Alver:2010gr} cannot be directly measured.
Other methods primarily rely on particle correlations and the concepts of Q-vectors and cumulants~\cite{Danielewicz:1985hn, Borghini:2000sa, Bilandzic:2010jr, Jia:2017hbm}.
An advantage of particle correlation methods is that they eliminate the need for event planes in Eq.~\eqref{oneParDis}.
In addition, cumulants can be concisely expressed through generating functions~\cite{Borghini:2000sa, Borghini:2001vi}.
This category of methods includes particle cumulants~\cite{Borghini:2000sa, Borghini:2001vi, Bilandzic:2010jr}, Lee-Yang zeros~\cite{Bhalerao:2003xf, Bhalerao:2003yq, FOPI:2005ukb}, and symmetric cumulants~\cite{Bilandzic:2013kga}, among other recent generalizations~\cite{Bhalerao:2013ina, DiFrancesco:2016srj, Mordasini:2019hut}.
More recently, some of us suggested the possibility of analyzing the flow coefficient via the maximum likelihood estimator (MLE)~\cite{sph-vn-10}.
The proposal is based on an established statistical inference approach~\cite{book-statistical-inference-Wasserman}, which views the flow coefficients as the unknown parameters of a hypothetical probability distribution when the experimental data are given.
A specifically chosen recipe for such estimation is known as an {\it estimator}, and the MLE is a well-known approach achieved by maximizing the associated likelihood function in the parameter space.
In other words, in our case the estimated parameters, namely, the flow harmonics, ensure that the observed data are the most probable.

In particular, the proposed method is advantageous with respect to two specific aspects of flow analysis.
The first is nonflow, a term referring to collective behavior that cannot be explained by independent emission, as per the one-particle distribution function Eq.~\eqref{oneParDis}.
Conservation laws such as momentum conservation might lead to deviations in particle correlations compared to cases where the particle spectrum is entirely flow-driven~\cite{Chajecki:2008yi}.
It is generally understood that the impact of nonflow diminishes as particle correlations involve more hadrons~\cite{Borghini:2001vi, Borghini:2001zr} or when a rapidity gap is introduced between particles~\cite{PHENIX:2003qra, Voloshin:2006wi, Voloshin:2006gz}.
The second aspect pertains to the finite multiplicity in individual events.
This factor introduces statistical uncertainty even if hadrons are emitted independently according to the one-particle distribution function~\cite{sph-vn-09, sph-corr-32}.
This uncertainty is distinct from those caused by initial condition fluctuations, which usually have a physical origin in the underlying microscopic model.
From a hydrodynamic perspective, the latter manifests as fluctuations in the initial energy distribution's geometry on an event-by-event basis.
Notably, the statistical error of the particle correlation method can be even more significant than that of the event plane method~\cite{Voloshin:2008dg}.
Furthermore, for events with lower multiplicity, multi-particle correlations are likely to be affected by greater statistical uncertainty~\cite{sph-corr-31}.
Precisely, for a given estimator of a specific observable, such uncertainty might be estimated analytically and be sizable~\cite{sph-vn-10}.
Because the MLE is an asymptotically normal estimator, its efficiency is guaranteed in the sense that it is more accurate than any other estimator at the limit of a significant sample size.
Moreover, as it is more flexible to introduce additional ansatz in the flow distribution function, it is arguably more potent to deal with possible scenarios with the presence of nonflow.
In terms of convergence, the method is either unbiased or asymptotically unbiased. 
The context of relativistic heavy-ion collisions meets most of the characteristics of MLE. 
Specifically, the measurements performed at RHIC and LHC have accumulated significant events for different collision systems at different centralities.

Besides flow harmonics, one may further generalize the discussions to quantities constructed by generic multi-particle correlators.
As discussed in~\cite{ALICE:2011svq, Gardim:2012im}, compared with flow harmonics, some quantities might be more sensitive to the specific characteristic of the underlying strongly interacting system.
Its deviation from the case of purely statistically independent emission measures the viability of the event planes as a well-defined quantity.
In particular, it has been argued that event-by-event fluctuations in the initial state might introduce a sizable effect in the multi-particle correlator, giving rise to $p_{\rm T}$ dependent event planes~\cite{Heinz:2013bua}.
To be more specific, the event planes in the final state can be estimated by using the particles' azimuthal distribution over a relatively broad transverse momentum range on an event-by-event basis.
However, both experimental data~\cite{ALICE:2022dtx, Zhou:2014bba} and hydrodynamic calculations~\cite{Alver:2010gr, Heinz:2013bua} indicate that the event planes of final-state particles fluctuate across different $p_{\rm T}$ ranges.
In particular, in~\cite{CMS:2015xmx, ALICE:2017lyf, Barbosa:2021ccw}, it was observed that the correlation matrix exhibits an approximate factorization in transverse momentum, which is considered favorable evidence for the hydrodynamic picture.
The breakdown of factorization is attributed mainly to event-by-event fluctuations in the initial energy distribution~\cite{ALICE:2011svq, Gardim:2012im, Heinz:2013bua, CMS:2015xmx, ALICE:2017lyf} rather than to the transport properties of the medium, indicating that valuable information on initial conditions can be extracted from such analysis.
However, as will be elaborated below, when evaluated in terms of the Q-vectors or flow vectors, there are some ambiguities regarding flow estimation by specific choices of the order of the multi-particle correlators.
On the other hand, these quantities can always be estimated in terms of the MLE, which is irrelevant to the specific choice of the correlators.

Motivated by the above considerations, the present study further applies the proposed method to analyze differential flow and multi-particle correlators.
We focus on collisions of Au+Au at 200 GeV and Pb+Pb at 2.76 TeV, where the differential flows obtained for varied centrality windows and momentum cuts have shown consistency when compared with conventional methodologies, such as particle correlation and event-plane methods.
We proceed to analyze the flow harmonics and event-plane correlations. 
Moreover, we show that MLE's unique capabilities allow us to compute specific correlators that are typically inaccessible by other means. 

The remainder of this paper is organized as follows. 
In the next section, we briefly review flow analysis methods, particularly the mathematical framework of the MLE and its application to flow analysis in relativistic heavy-ion collisions. 
In Sec.~\ref{section3}, we conduct numerical studies to evaluate the differential flow and event planes using different approaches and compare the results to the data and those obtained using simulations.
In Sec.~\ref{section4}, we apply the method to analyze the flow factorization and event plane correlations.
The results are again compared with those obtained using multi-particle cumulants and event-plane methods.
Last but not least, in Sec.~\ref{section5}, we elaborate on high-order and mixed harmonic factorization breakdown, particularly in the cases that are not straightforward for the particle-correlation method unless the event planes are perfectly correlated.
The last section is devoted to further discussions and concluding remarks.

\section{Statistical estimators for flow harmonics}\label{section2}

As mentioned, the most prominent methods for extracting flow harmonics are based on particle correlations~\cite{Poskanzer:1998yz}.
The foundation of these methods relies on the following relation for $k$-particle correlations~\cite{Bhalerao:2011yg}:
\begin{eqnarray}
\langle k\rangle_{n_1,\cdots,n_k}\equiv \langle e^{i(n_1\phi_1 + \cdots + n_k\phi_k)} \rangle
= v_{n_1}\cdots v_{n_k} e^{i\left(n_1\Psi_{n_1}+\cdots+n_k\Psi_{n_k}\right)} , \label{nkCorr}
\end{eqnarray}
where $\langle\cdots\rangle$ denotes an average over distinct tuples of particles, assuming independent particle emission as described by Eq.~\eqref{oneParDis} in the limit of infinite multiplicity.

To focus on $v_n$, one typically selects a specific set of $(n_1, \cdots, n_k)$, such that~\cite{Bhalerao:2011yg}
\begin{eqnarray}
\sum_{j=1}^k n_j=0 ,\lb{sumRes}
\end{eqnarray}
which ensures that all coefficients involving the event planes cancel out in the exponential.
Effectively, the event-plane correlator becomes part of the formalism~\cite{Bhalerao:2013ina}.
The simplest example is the two-particle correlations ($k=2$), where one takes $n_1 = -n_2 = n$.
Subsequently, we have
\begin{eqnarray}
\langle 2 \rangle_{n,-n} \equiv \left \langle e^{in(\phi_{1}-\phi_{2} )}\right \rangle
= \langle\cos{n(\phi_{1}-\phi_{2})} \rangle = v_n^2
\label{eq2}
\end{eqnarray}
In practice, however, the multiplicity $M$ is a finite number for a realistic event. 
The analysis is carried out given the discrete values of the azimuthal angles of the measured particles, namely, $\phi_{1}, \phi_{2}, \cdots, \phi_{M}$.
Instead of an integration implicitly performed in Eq.~\eqref{eq2}, it is intuitive to use the following summation~\cite{Bilandzic:2013kga}
\begin{eqnarray}
\widehat{v_n^2} =\frac{1}{M(M-1)}\sum_{i\ne j}\cos n(\phi_{i}-\phi_{j}) ,
\label{eqEst2}
\end{eqnarray}
to assess the flow harmonics $v_n$.
In statistical inference, Eq.~\eqref{eqEst2} is an {\it estimator}.
It furnishes an estimation for $v_n^2$ instead of $v_n$, based on a finite number of measurements.
The first two moments of this estimator can be readily evaluated~\cite{Bilandzic:2013kga}.
Although an unbiased estimator, it possesses a finite variance that decreases as the multiplicity increases.
For higher order correlators, one can either employ the generating function~\cite{Borghini:2000sa, Borghini:2003ur, Borghini:2007ku} or evaluate them directly~\cite{Bilandzic:2010jr} using the $Q$-vectors~\cite{Danielewicz:1985hn}.
As a generalization of the unweighted summand in Eq.~\eqref{nkCorr}, quantities constructed by the $Q$-vectors can also be viewed as estimators.
Moreover, it is straightforward to demonstrate that the variance of these quantities remains finite~\cite{sph-corr-31}.
A variance of finite size implies uncertainty subjected to limited statistics, particularly the finite number of events.

Following this line of thought, it can be argued that MLE could serve as an alternative estimator for flow and related observables, a possibility explored in a previous study~\cite{sph-vn-10}.
Expressly, for a given set of observations $y\equiv (y_1, y_2, \cdots, y_M)$, we assume that they are sampled from a joint probability distribution governed by several unknown parameters $\theta \equiv (\theta_1, \theta_2, \cdots, \theta_m)$.
As mentioned in the Introduction, the likelihood function $\mathcal{L}$ for the observed data is given by:
\begin{eqnarray}
\mathcal{L}(\theta) \equiv \mathcal{L}(\theta; y) = f(y; \theta) ,\label{defLn}
\end{eqnarray}
which represents the joint probability density for the given observations evaluated for the parameters $\theta$.
The goal of MLE is to determine the parameters for which the observed data attains the highest joint probability, namely:
\begin{eqnarray}
\hat{\theta}_{\mathrm{MLE}}=\arg\max\limits_{\theta\in\Theta}\mathcal{L}(\theta) , \label{defMLE}
\end{eqnarray}
where $\Theta$ is the domain of the parameters.
In particular, for independent and identically distributed (i.i.d.) random variables, $f(y; \theta)$ is given by the product of individual distributions $f^\mathrm{uni}$:
\begin{eqnarray}
f(y; \theta)=\prod_{j=1}^M f^\mathrm{uni}(y_j; \theta) . \label{iidfn1}
\end{eqnarray}

This scheme can be readily applied to collective flow in heavy-ion collisions.
Considering an event consisting of $M$ particles, the likelihood function reads:
\begin{eqnarray}
\mathcal{L}(\theta; \phi_{1}, \cdots, \phi_M) = f(\phi_{1},\cdots, \phi_M; \theta)=\prod_{j=1}^{M}f_1(\phi_j; \theta) ,
\label{eqlikelihood}
\end{eqnarray}
where the likelihood function $\mathcal{L}$ is governed by the one-particle distribution function Eq.~\eqref{oneParDis}.
The last equality is based on the assumption that the particles' azimuthal angles are i.i.d. variables.
The parameters of the distribution, $\theta = (v_1, \Psi_1, v_2, \Psi_2, \cdots)$, are the flow harmonics and the event planes.

In practice, one often chooses the objective function to be the log-likelihood function ${\ell}$:
\begin{eqnarray}
{\ell}(\theta; \phi_{1}, \cdots, \phi_M) = \log\mathcal{L}(\theta; \phi_{1}, \cdots, \phi_M)
= \sum_{j=1}^{M}\log f_1(\phi_j; \theta) .
\label{eqlogl}
\end{eqnarray}
Numerical calculations indicate that Eq.~\eqref{eqlogl} is more favorable than Eq.~\eqref{eqlikelihood}, although as multiplicity $M$ increases, an appropriate implementation should be adopted to avoid the increasing truncation error.

The maximum of $\ell$ occurs at the same value of $\theta$, which maximizes $\mathcal{L}$.
For $\ell$ that is differentiable in its domain $\Theta$, the necessary conditions for the occurrence of a maximum are:
\begin{eqnarray}
\frac{\partial{\ell}}{\partial\theta_1}=\cdots=\frac{\partial{\ell}}{\partial\theta_m}=0 .
\label{condMLE}
\end{eqnarray}

As discussed in the Introduction, MLE has asymptotic normality, which attains the Cramér-Rao lower bound as the sample size increases.
In other words, no consistent estimator has a lower asymptotic mean squared error than the MLE.
In the context of relativistic heavy-ion collisions, all events of a given multiplicity asymptotically form a (multivariate) normal distribution:
\begin{eqnarray}
\hat{\theta}_{\mathrm{MLE}} \sim N\left(\theta_0, (I_M(\theta_0))^{-1}\right) ,
\label{assNormMLE}
\end{eqnarray}
where $\theta_0$ represents the true value, and $I_M(\theta)$ is the Fisher information matrix, defined as:
\begin{eqnarray}
I_M(\theta) \equiv E_\theta\left[-\frac{d^2}{d\theta^2}{\ell}(\theta;\phi_1,\cdots,\phi_M)\right] ,
\label{defIM}
\end{eqnarray}
where the expectation $E_\theta$ is taken with respect to the distribution $f(\phi_1,\cdots,\phi_M;\theta)$.
For i.i.d. data, the Fisher information possesses the form:
\begin{eqnarray}
I_M(\theta) = M I_1(\theta),
\label{defI1}
\end{eqnarray}
where $I_1$ is the Fisher information matrix for a single observation.
As a result, the standard deviation of MLE is expected to be roughly proportional to $\frac{1}{\sqrt{M}}$.
As discussed in~\cite{sph-vn-10}, these properties can be further quantified using the Wald, likelihood ratio, and score tests.

\section{Differential flow and event planes using the MLE method}\label{section3}

In a previous study~\cite{sph-vn-10}, MLE was employed as an estimator of the integrated flow. 
The present work first generalizes the approach for evaluating the differential flow.
Compared with integrated flow, evaluating differential flow requires more computational power and is potentially limited by statistics.
This is because MLE involves evaluating partial derivatives of the likelihood, which becomes increasingly complicated as the number of particles increases.
Nonetheless, we show that such an implementation is indeed feasible.

Specifically, we apply the MLE method to evaluate the harmonic coefficients as functions of the transverse momentum.
For a given momentum interval, the parameters $\theta$ of the underlying distribution function are taken to be the flow harmonics and event planes $v_n$ and $\Psi_n$, where we consider the index up to $n=4$.
Numerical simulations, as well as experimental data, are employed for the analysis.
The simulated data are generated on an event-by-event fluctuating basis using the hydrodynamic code NeXSPheRIO~\cite{sph-review-01, sph-review-02}.
All the evaluated quantities involve all charged particles.

\begin{figure}[ht]
\centerline{\includegraphics[height=0.7\textwidth]{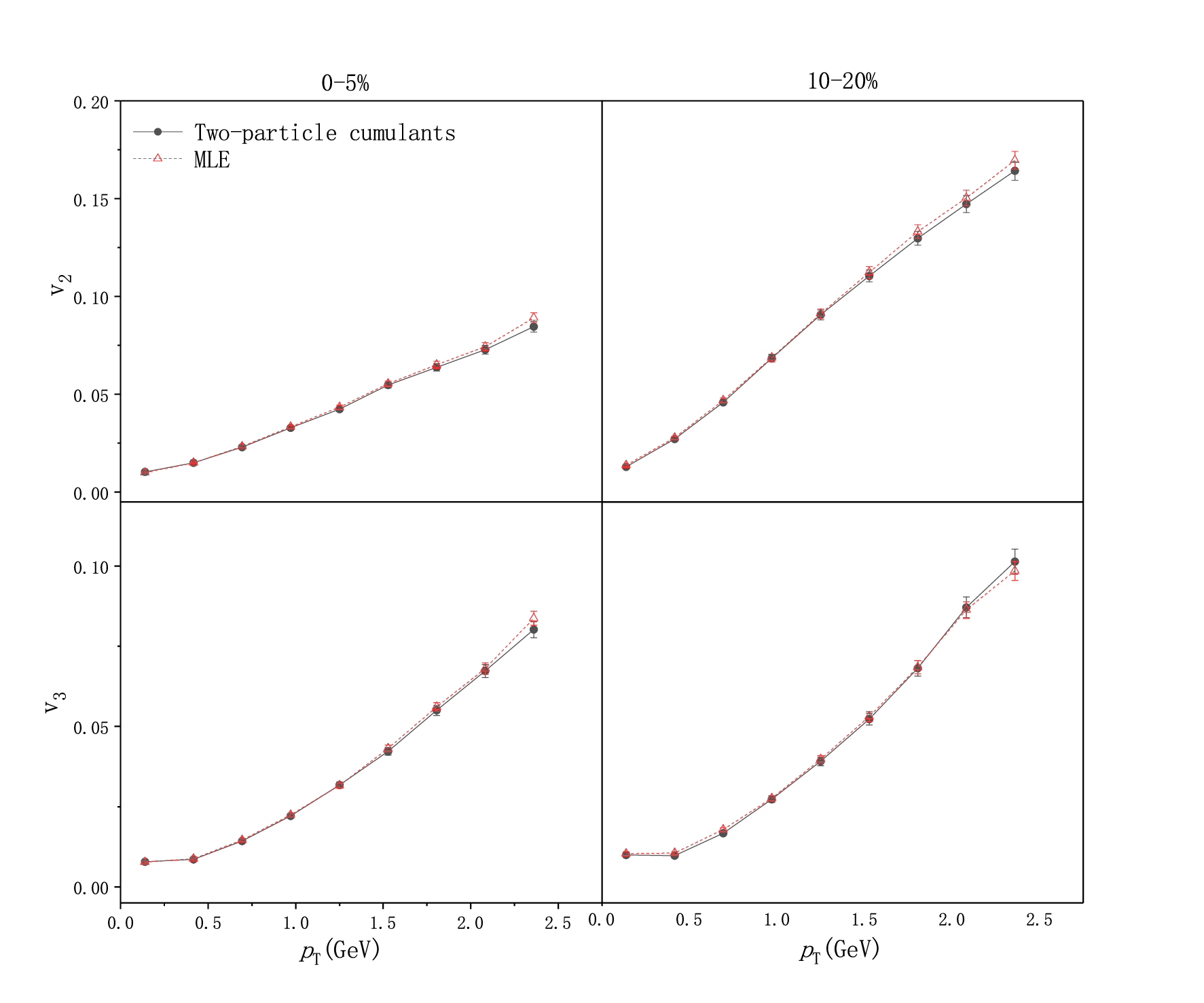}}
\renewcommand{\figurename}{Fig.}
\vspace{-0.5cm}
\caption{The elliptic and triangular differential flows for the 0-5\% and 10-20\% centrality windows of Au+Au collisions at 200 GeV.
The calculations are based on the hydrodynamic simulations using the NeXSPheRIO code. 
The upper row presents the results for the elliptic flow $v_2$, while the lower row shows those for the triangular flow $v_3$.  
The left column shows the results for the centrality 0-5\%, and the right column represents that for 10-20\%.
The error bars represent the standard errors.}
\label{fig_differentialflow}
\end{figure}

\begin{figure}[ht]
\centerline{\includegraphics[height=0.4\textwidth]{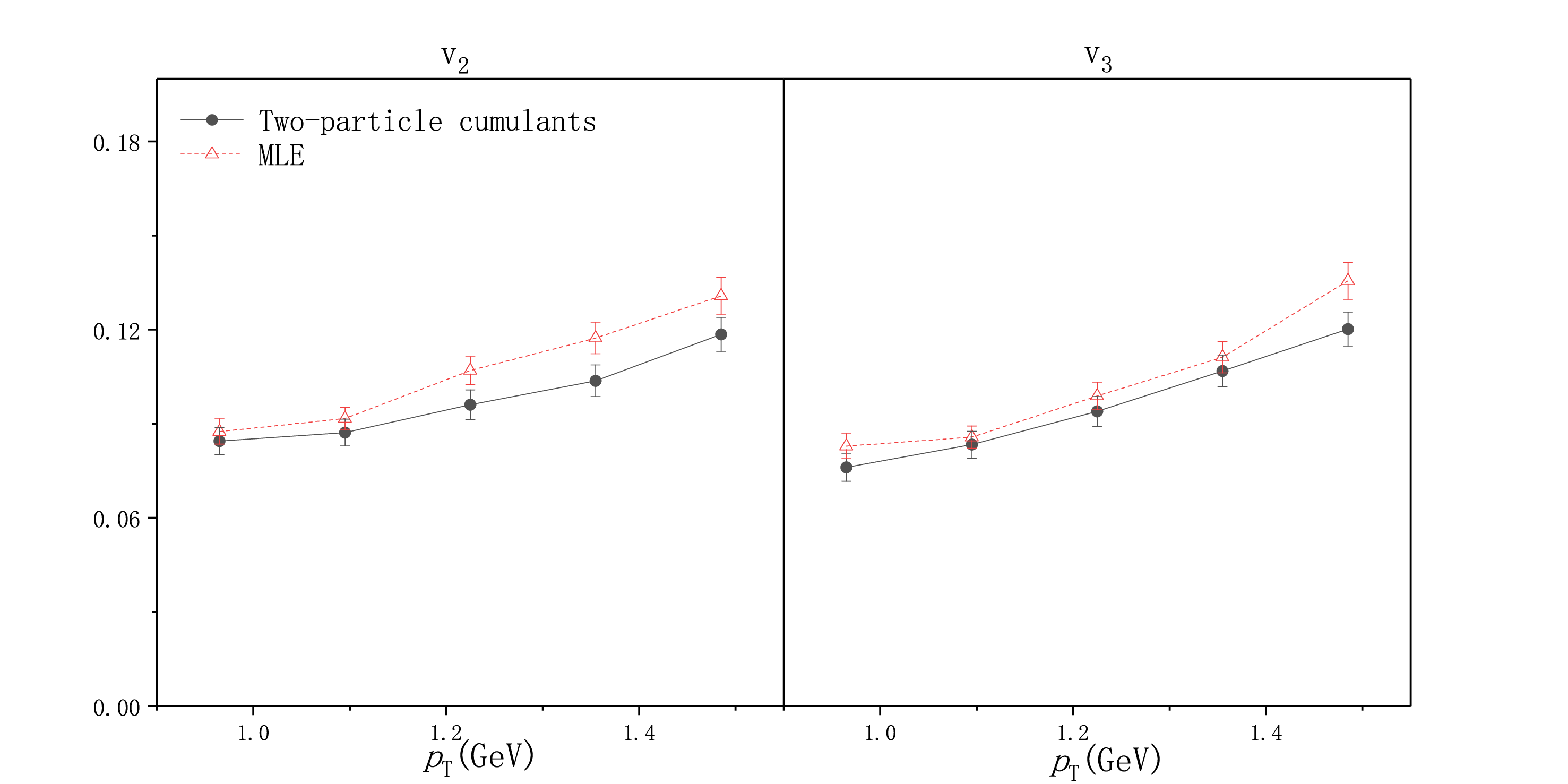}}
\renewcommand{\figurename}{Fig.}
\caption{The same as Fig.~\ref{fig_differentialflow} but for minimum bias events from Pb+Pb collisions at 2.76 TeV.
The calculations are performed using CMS Open Data.
}
\label{fig_differentialflow_CMS}
\end{figure}

In Fig.~\ref{fig_differentialflow}, we present the calculated differential flows, $v_2(p_{\rm T})$ and $v_3(p_{\rm T})$, as functions of transverse momentum for Au+Au collisions at $\sqrt{s_{{\rm NN}}}=$ 200 GeV, using different methods. 
Specifically, the results obtained with MLE are compared to those derived from the two-particle correlator in Eq.~\eqref{eq2}. 
The differential flows for the 0-5\% and 10-20\% centrality windows are obtained by averaging over 500 and 480 simulated events, respectively.
Generally, for different harmonics and centralities, the MLE results agree reasonably well with those using two-particle cumulants.
Minor deviations are observed at more significant transverse momenta.
Similar calculations are also carried out using CMS Open Data for Pb+Pb collisions at $\sqrt{s_{{\rm NN}}}=$ 2.76 TeV~\cite{LHC-cms-open-data-02} for minimum bias events.
In Fig.~\ref{fig_differentialflow_CMS}, the calculated differential flows are presented for the transverse momentum interval $0.9 < p_{\rm T} < 1.7$ GeV.
The choice of such an interval is due to a peculiar feature of the 2010 Pb+Pb data, where a considerably tighther tracking selection was needed in order to deal with fake tracks contamination.
Specifically, the data set contains no particle in the low-momentum region.
On the other hand, for higher transverse momentum $p_{\rm T}> 1.7$ GeV, the data also become relatively scarce.
For instance, out of 2956 events, only 720 contain more than 500 charged hadrons.
Although the discrepancies are more significant when compared to Fig.~\ref{fig_differentialflow}, the resulting differential flows evaluated using the two approaches primarily agree.
It should be noted that the differential flow obtained by particle correlation does not consider charged particle reconstruction and selection efficiency corrections as employed, e.g., in Ref.~\cite{CMS:2013jlh}.

\begin{table}
     \caption{Integrated flows and their standard deviations for various momentum intervals obtained using different methods.
     The values are essentially event-by-event averages of the flow harmonics evaluated using individual hadrons falling within the respective momentum intervals.}\label{tabvn1e}
     \begin{tabular}{cccccc}
          \hline\hline
          flow &~~~~momentum interval (GeV)~~&~~MLE~~~~&~~~~~~event-plane~~~~&  two-particle cumulant &  four-particle cumulant\\
          \hline
          $v_2$  &  $0.1<p_{\rm T}<0.5$  & 0.025 $\pm$ 0.020  &~~~~0.027 $\pm$ 0.014  & 0.027 $\pm$ 0.015  & 0.028 $\pm$ 0.009\\
          $v_3$  &                 & 0.024 $\pm$ 0.019  &~~~~0.025 $\pm$ 0.013  & 0.025 $\pm$ 0.013  & 0.028 $\pm$ 0.007\\
          \hline
          $v_2$  &  $0.5<p_{\rm T}<1.0$  & 0.034 $\pm$ 0.025  &~~~~0.034 $\pm$ 0.019  & 0.034 $\pm$ 0.019 & 0.034 $\pm$ 0.015\\
          $v_3$  &                 & 0.029 $\pm$ 0.021  &~~~~0.028 $\pm$ 0.015  & 0.028 $\pm$ 0.015 & 0.030 $\pm$ 0.009\\
          \hline
          $v_2$  &  $1.0<p_{\rm T}<1.5$  & 0.056 $\pm$ 0.043  &~~~~0.058 $\pm$ 0.030  & 0.058 $\pm$ 0.032 & 0.056 $\pm$ 0.025\\
          $v_3$  &                 & 0.048 $\pm$ 0.037  &~~~~0.049 $\pm$ 0.025  & 0.048 $\pm$ 0.022 & 0.048 $\pm$ 0.016\\
          \hline\hline
     \end{tabular}
\end{table}

\begin{figure}[ht]
\centerline{\includegraphics[height=0.7\textwidth]{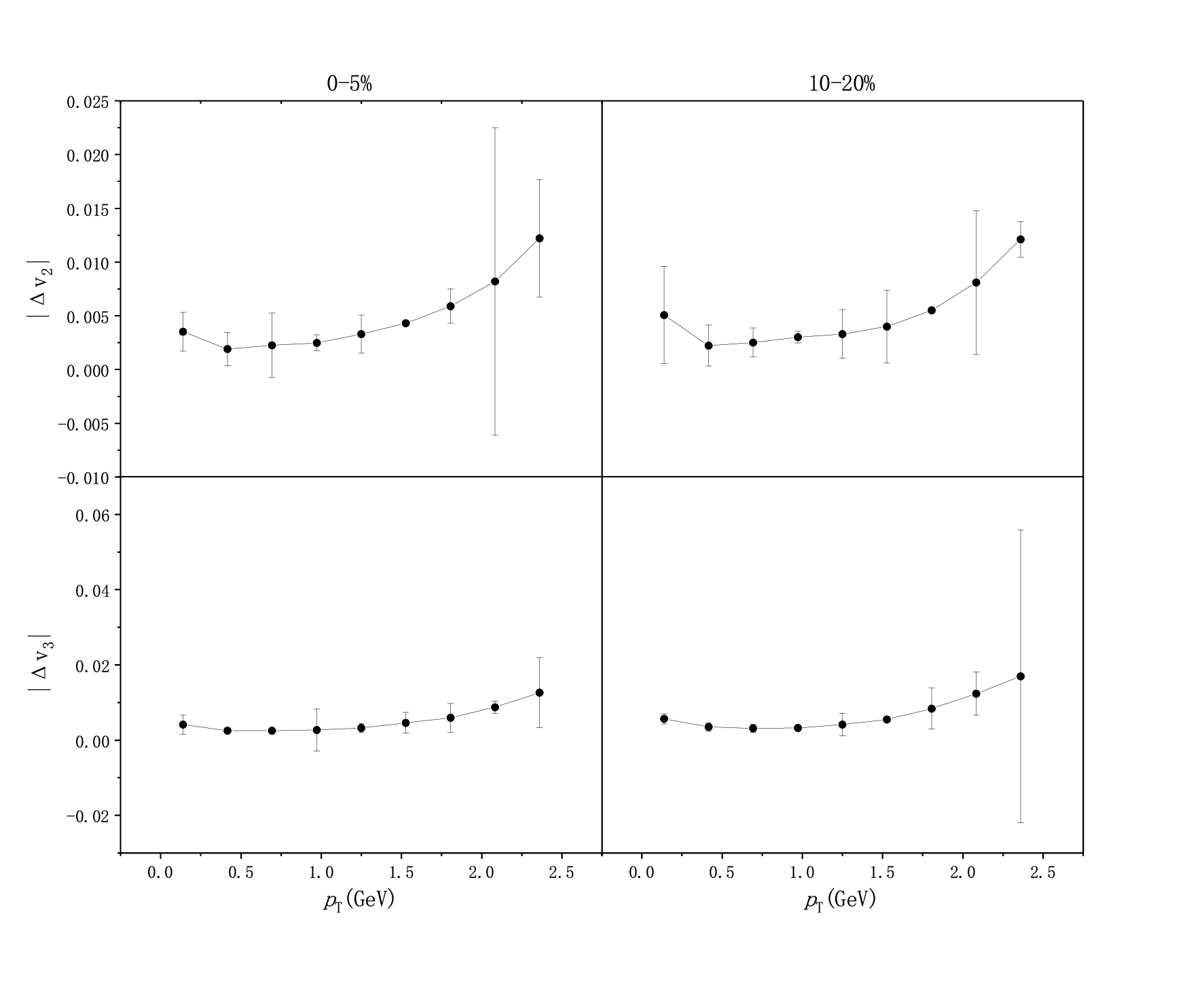}}
\renewcommand{\figurename}{Fig.}
\vspace{-0.5cm}
\caption{The difference of the calculated flow harmonics using the MLE and two-particle cumulant methods. 
The calculations are carried out for the 0-5\% and 10-20\% centrality windows of Au+Au collisions at 200 GeV, based on the hydrodynamic simulations using the NeXSPheRIO code. 
The upper row presents the results for $v_2$, while the lower row shows those for $v_3$.  
The left column shows the results for the centrality 0-5\%, and the right column represents that for 10-20\%.
The error bars indicate the standard deviations of the difference.
}
\label{fig_deviationVn}
\end{figure}

\begin{figure}[ht]
\centerline{\includegraphics[height=0.7\textwidth]{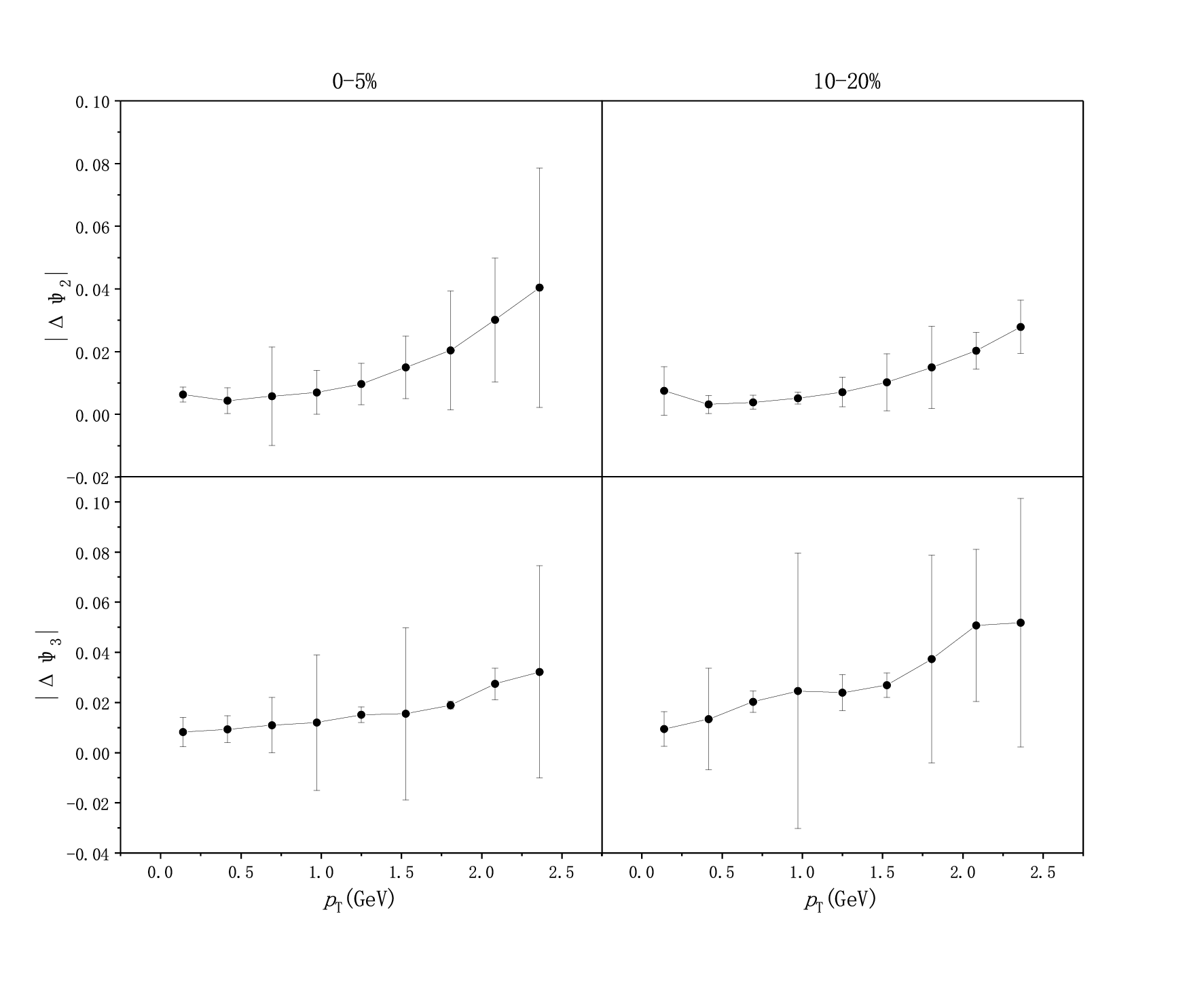}}
\renewcommand{\figurename}{Fig.}
\caption{Similar to Fig.~\ref{fig_deviationVn}, but for the difference of the calculated event planes using the MLE and event-plane methods.}
\label{fig_deviationEP}
\end{figure}

\begin{table}
\caption{The standard deviation of the elliptical flow and the corresponding event plane as functions of the number of events, extracted using the MLE method.}\label{tabvn1e2}
\begin{tabular}{c ccccc}
        \hline\hline
          number of events    &~~~~~~200~~~~~~&~~~~~~500~~~~~~&~~~~~~700~~~~~~&~~~~~~1500~~~~~~&~~~~~~1900~~~~~~\\
          \hline
          $\sigma_{v_2}$      &  0.0155     & 0.0156       & 0.0155  &  0.0155    & 0.0154 \\
          \hline
         $\sigma_{\Psi_2}$    &  0.842   &  0.899   & 0.901 &  0.905   &  0.899  \\
         \hline\hline
\end{tabular}
\end{table}

We proceed further to quantitatively examine the difference in flow harmonics and event planes across different flow analysis methods.
Besides those seen in Fig.~\ref{fig_differentialflow}, we show an explicit comparison of the integrated flow harmonics obtained using different estimators.
In Tab.~\ref{tabvn1e}, the integrated flows are evaluated using four methods: MLE, event-plane method, two-particle cumulant, and four-particle cumulant. 
The results are obtained by averaging over 1900 events within the 0-5\% centrality window.
Because of their different definitions, some difference is observed between the flow harmonics obtained using different methods.
In particular, the flow harmonics obtained using two-particle cumulant are evaluated by $\sqrt{\widehat{v_n^2}}$, as given by Eq.~\eqref{eq2}.
As a comparison, the MLE estimates the harmonic coefficient itself, as a parameter of the underlying statistical distribution that generates the observed particle spectrum with the largest probability.
As an asymptotically normal estimator, MLE's deviation from the true value essentially comes from its bias at lower statistics.
While $\langle 2 \rangle_{n,-n}$ is not a biased estimator for $v_n^2$, by definition, $v_n\{2\}$ gives a biased estimation for $v_n$, attributed to a skewed distribution associated with flow fluctuations~\cite{hydro-corr-ph-09}.
Therefore, at more significant statistics, $v_n\{2\}$ tends to overestimate the true value as observed in Tab.~\ref{tabvn1e}.
Nonetheless, the difference observed in differential flow for these two approaches is not significant.
As for the differential flow, in Figs.~\ref{fig_deviationVn} and~\ref{fig_deviationEP}, we show the difference in the obtained flow harmonics between the MLE and the two-particle cumulant methods, as well as that for event planes between the MLE and the event-plane methods.
In Tab.~\ref{tabvn1e2}, the event-average standard deviations of the estimated elliptic flow and the corresponding event plane are presented.
The standard deviations of flow harmonics are found to be much smaller than those for the event planes, and the tendency to converge is observed for both quantities as the number of events increases\footnote{
Conversely, it is noted that the values and uncertainties of $v_2\{2\}$ and $v_2\{4\}$ shown in Tab.~\ref{tabvn1e} obtained using particle cumulants have not yet convergent to the statistical limits ${v_2\{2\}}>{v_2\{4\}}$~\cite{hydro-corr-ph-09} and $\sigma_{v_2\{2\}} < \sigma_{v_2\{4\}}$~\cite{hydro-corr-ph-04}.
We understood that this is due to that the number of events used to generate these quantities might not be sufficient.}.
While the magnitude and overall shape of the differential flows obtained using different approaches mostly agree, it is observed that the difference and their standard deviations in flow harmonics are smaller compared to those in event planes.
In other words, for a given event, it is more difficult to precisely estimate the event plane because of the sizable uncertainties, manifested as {\it parameter degeneracy} in the context of Bayesian inference, and the difference between different methods. 
Besides, although not explicitly shown here, the fluctuations of event planes evaluated at different momenta are rather significant.
Together with event-by-event flow correlations, this information constitutes the factorization breakdown, which will be further explored in the following section.

\section{Flow factorization and event-plane correlation}\label{section4}

In this section, we further explore flow factorization and event-plane correlations.
Flow factorization~\cite{Gardim:2012im, Heinz:2013bua} is understood to serve as a sensitive tool for probing the properties of the initial state fluctuations.
Specifically, it is measured by a Pearson correlation in terms of flow vectors $V_{n\Delta}$ evaluated at different transverse momenta: 
\begin{eqnarray}
       r_n(p_{\rm T}^{\rm a},p_{\rm T}^{\rm t})=\frac{V_{n\Delta}(p_{\rm T}^{\rm a},p_{\rm T}^{\rm t})}{\sqrt{V_{n\Delta}(p_{\rm T}^{\rm a},p_{\rm T}^{\rm a})V_{n\Delta}(p_{\rm T}^{\rm t},p_{\rm T}^{\rm t})}} ,
\label{rn_fact}
\end{eqnarray}
where $p_{\rm T}^{\rm t}$ and $p_{\rm T}^{\rm a}$ are transverse momenta of the trigger and associated particles, and $V_{n\Delta}(p_{\rm T1},p_{\rm T2})$ is the $n$th harmonic of the underlying di-hadron azimuthal distribution with transverse momenta $p_{\rm T1}$ and $p_{\rm T2}$, namely,
\begin{eqnarray}
       V_{n\Delta}(p_{\rm T1},p_{\rm T2})
       \equiv \left\langle e^{in\left(\phi(p_{\rm T1})-\phi(p_{\rm T2})\right)}\right\rangle 
       =\langle \cos n\left(\phi(p_{\rm T1}) -\phi(p_{\rm T2})\right) \rangle 
       =\langle V_n^*(p_{\rm T1})V_n(p_{\rm T2})\rangle ,
       \label{DefVnDelta}
\end{eqnarray}
where 
\bqn
V_n(p_{\rm T})=v_n(p_{\rm T}) e^{-in\Psi_n(p_{\rm T})},
\eqn 
is known as the flow harmonic vector~\cite{Luzum:2013yya, Ollitrault:2012cm}. 
Because of its explicit consideration of transverse momentum dependence, Eq.~\eqref{DefVnDelta} can be viewed as the {\it differential} counterpart of the two-particle correlation defined in Eq.~\eqref{eq2}.

By assuming that particle emission is independent and employing the specific form of one-particle distribution Eq.~\eqref{oneParDis}, the above ratio can be rewritten as
\begin{eqnarray}
       r_n(p_{\rm T}^{\rm a},p_{\rm T}^{\rm t})
       =\frac{\langle V_n^*(p_{\rm T}^{\rm a})V_n(p_{\rm T}^{\rm t})\rangle}{\sqrt{\langle V_n^*(p_{\rm T}^{\rm a})V_n(p_{\rm T}^{\rm a}) \rangle \langle V_n^*(p_{\rm T}^{\rm t})V_n(p_{\rm T}^{\rm t}) \rangle}}
       =\frac{\langle v_n(p_{\rm T}^{\rm a})v_n(p_{\rm T}^{\rm t}) \cos{n(\Psi_n(p_{\rm T}^{\rm a})-\Psi_n(p_{\rm T}^{\rm t}))} \rangle}{\sqrt{\langle v_n^2(p_{\rm T}^{\rm a}) \rangle \langle v_n^2(p_{\rm T}^{\rm t}) \rangle}}.
\label{rn_fact_qvec}
\end{eqnarray}
By definition, such a ratio would be unity if one assumes that the event planes $\Psi_n$ are global constants and meanwhile, the flow harmonics $v_n$ do not fluctuate at all, or their values evaluated at different momenta are perfectly correlated.

As discussed above, these quantities are understood to carry more detailed information on the initial state fluctuations and the viability of the event planes as a well-defined quantity.
On the one hand, by Eq.~\eqref{rn_fact}, the factorization ratio can be calculated using the di-hadron correlations, reminiscent of the estimator Eq.~\eqref{eqEst2}.
In practice, this particle correlation estimator is the most utilized in literature.
On the other hand, in terms of Eq.~\eqref{rn_fact_qvec}, one can also estimate the flow harmonics and event plane independently using the MLE method at different transverse momenta and evaluate the resulting factorization ratio on an event-by-event basis.
The latter potentially furnishes further insights into how flow and event-plane correlations contribute individually to the factorization ratio, as discussed below.

\begin{figure}[ht]
\centerline{\includegraphics[height=0.8\textwidth]{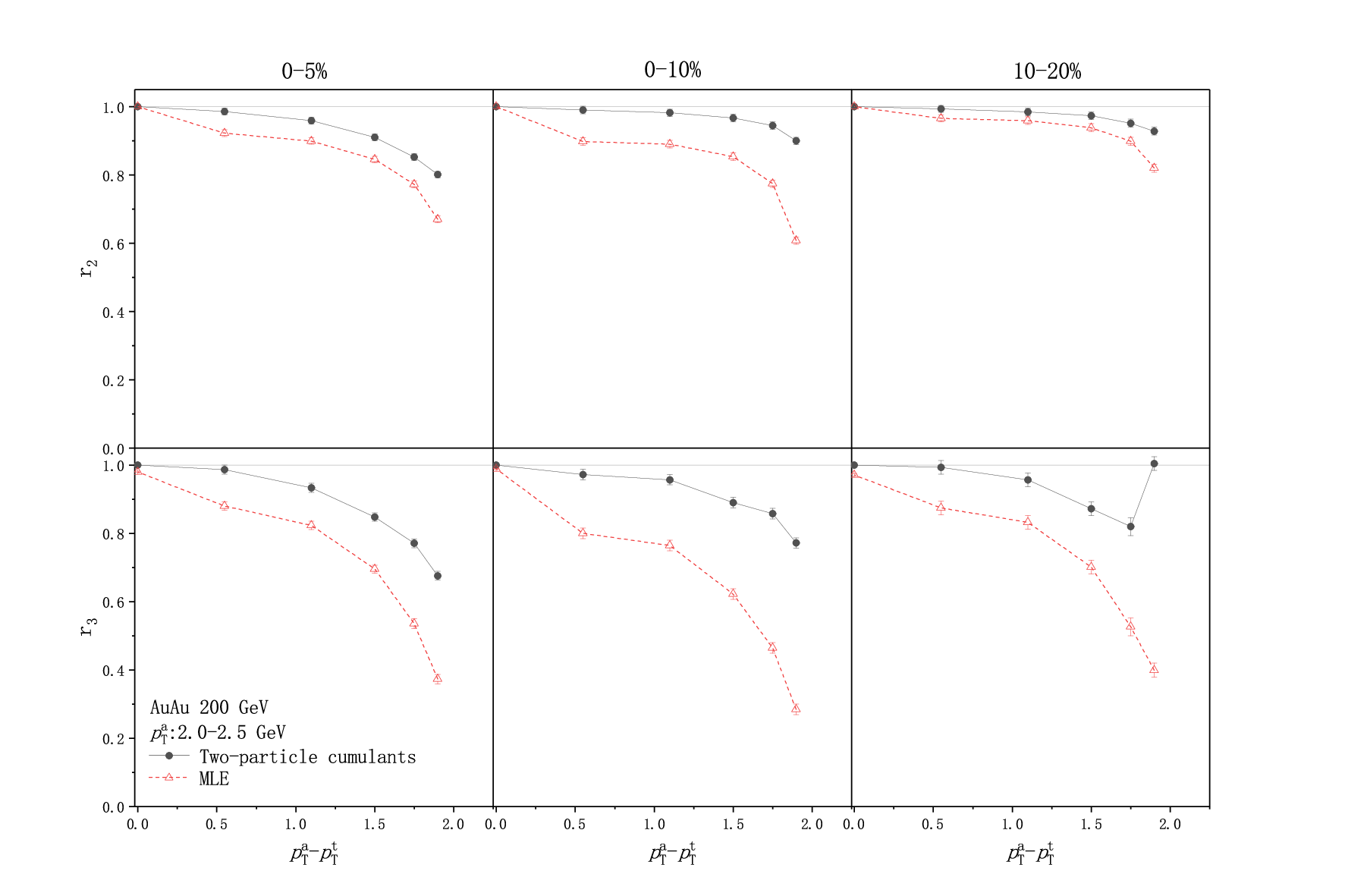}}
\renewcommand{\figurename}{Fig.}
\caption{The obtained factorization $r_n$ as a function of $p^{\rm a}_{\rm T}-p^{\rm t}_{\rm T}$ for Au+Au collisions at 200 GeV.
The calculations are based on numerical simulations using the NeXSPheRIO hydrodynamic code.
From left to right, the three columns correspond to three different centrality windows: 0-5\%, 0-10\%, and 10-20\%.
The upper row shows the results for $r_2$, while the lower row displays the calculated $r_3$.
For this figure and the ones below, the error bars represent the standard errors.
}
\label{fig_ecntrality}
\end{figure}

\begin{figure}[ht]
\centerline{\includegraphics[height=0.8\textwidth]{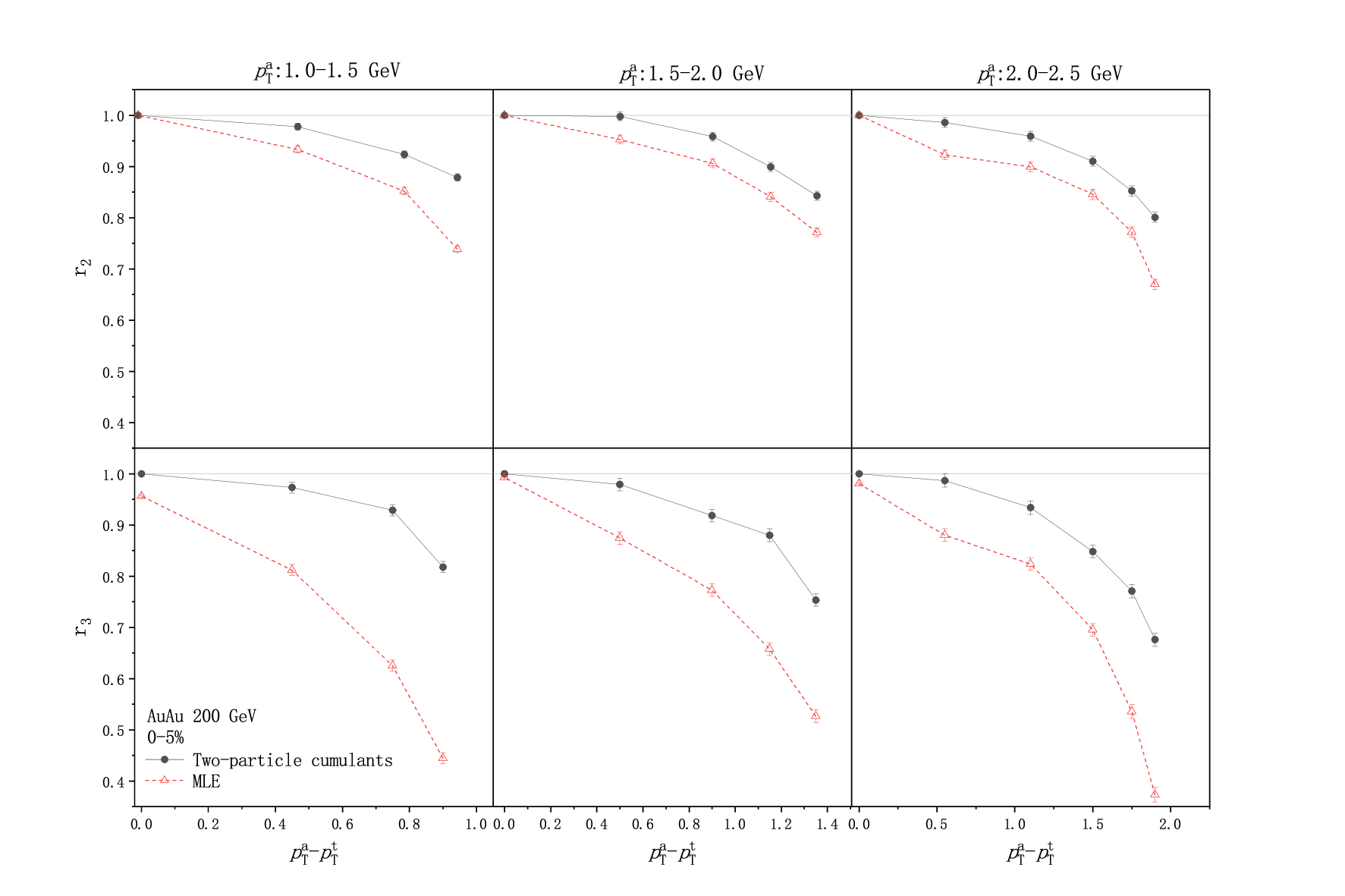}}
\renewcommand{\figurename}{Fig.}
\caption{The obtained factorization $r_n$ as a function of $p^{\rm a}_{\rm T}-p^{\rm t}_{\rm T}$ for 0-5\% centrality for different $p_\mathrm{T}^a$ ranges for Au+Au collisions at 200 GeV.
The calculations are based on numerical simulations using the NeXSPheRIO hydrodynamic code.
From left to right, the three columns correspond to three different $p^{\rm a}_{\rm T}$ interval selections.
The upper row shows the results for $r_2$, while the lower row displays the calculated $r_3$.
}
\label{rn510}
\end{figure}

\begin{figure}[ht]
\centerline{\includegraphics[height=0.8\textwidth]{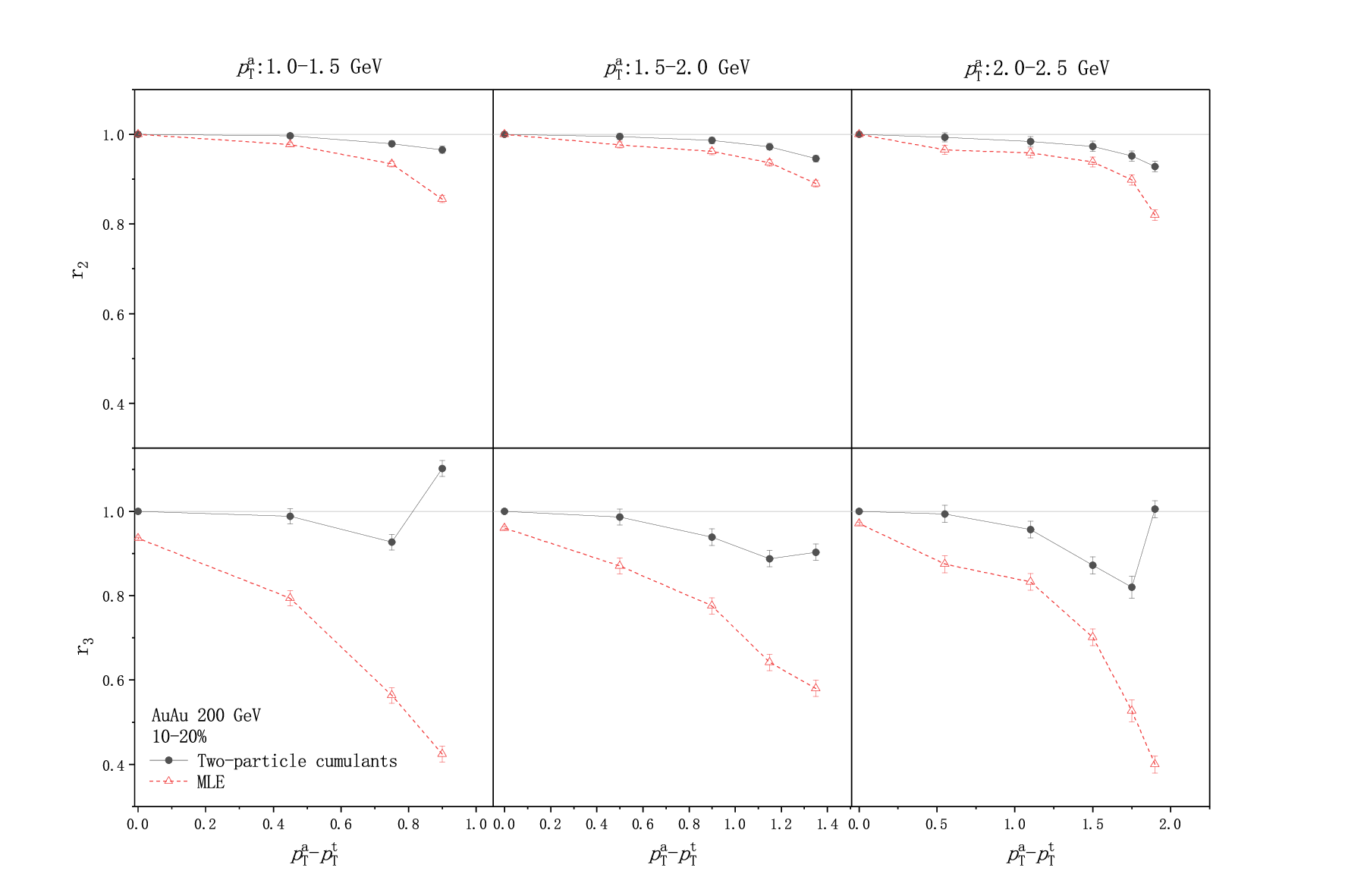}}
\renewcommand{\figurename}{Fig.}
\caption{The same as Fig.~\ref{rn510} but for 10-20\% centrality window.}
\label{rn1020}
\end{figure}

\begin{figure}[ht]
\centerline{\includegraphics[height=0.8\textwidth]{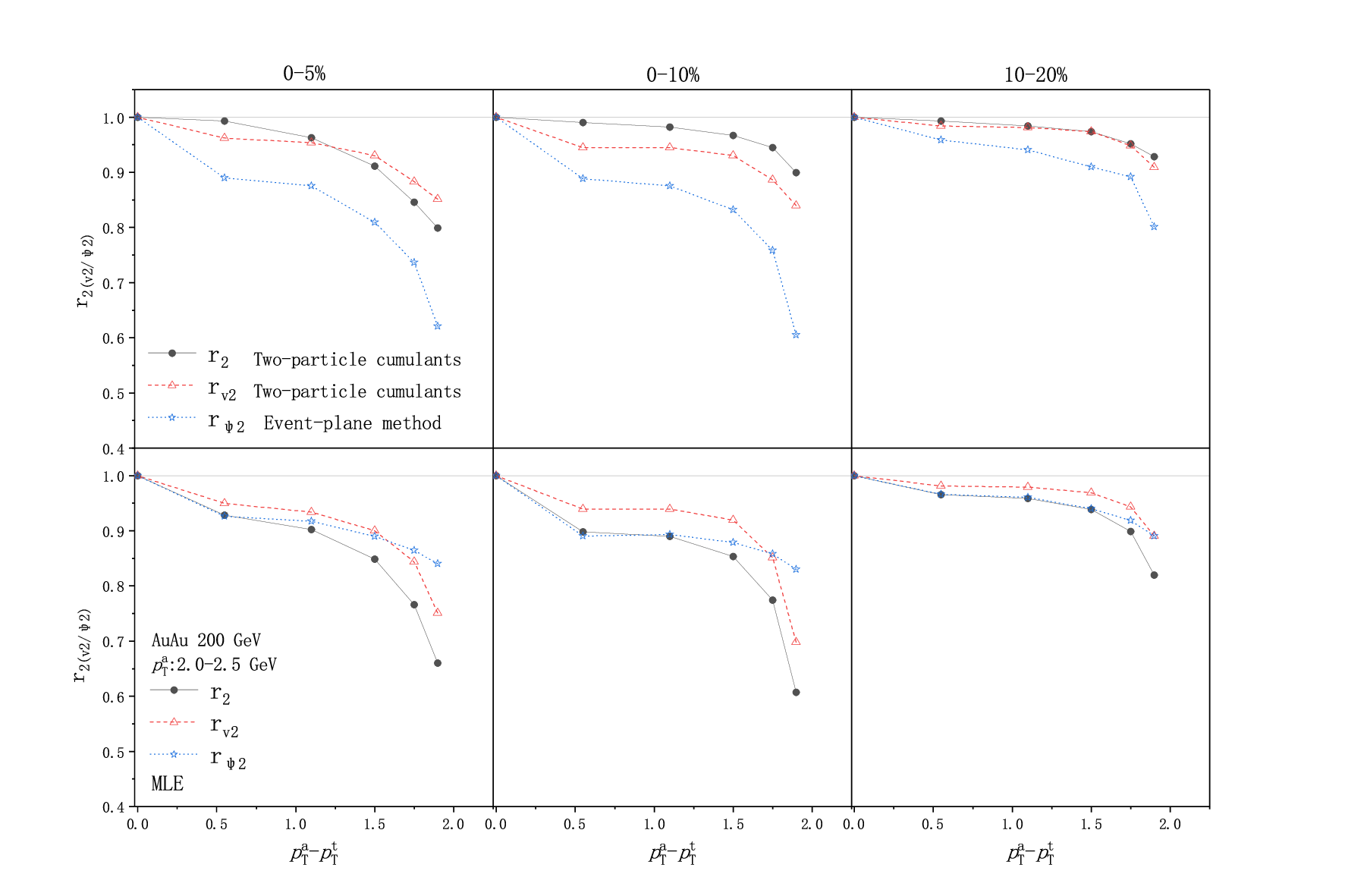}}
\renewcommand{\figurename}{Fig.}
\renewcommand{\figurename}{Fig.}
\caption{The obtained factorizations $r_2$, $r_{v_2}$, and $r_{\Psi_2}$ as functions of $p^{\rm a}_{\rm T}-p^{\rm t}_{\rm T}$ for the Au+Au collisions at 200 GeV.
The calculations are based on numerical simulations using the NeXSPheRIO hydrodynamic code.
From left to right, the three columns correspond to three different centrality windows: 0-5\%, 0-10\%, and 10-20\% with $2.0 < p_\mathrm{T}^a <2.5$ GeV.
The upper row presents the results for the two-particle cumulants and event-plane method, while the lower row shows those for MLE.
}
\label{MLECORR}
\end{figure}

\begin{figure}[ht]
\centerline{\includegraphics[height=0.8\textwidth]{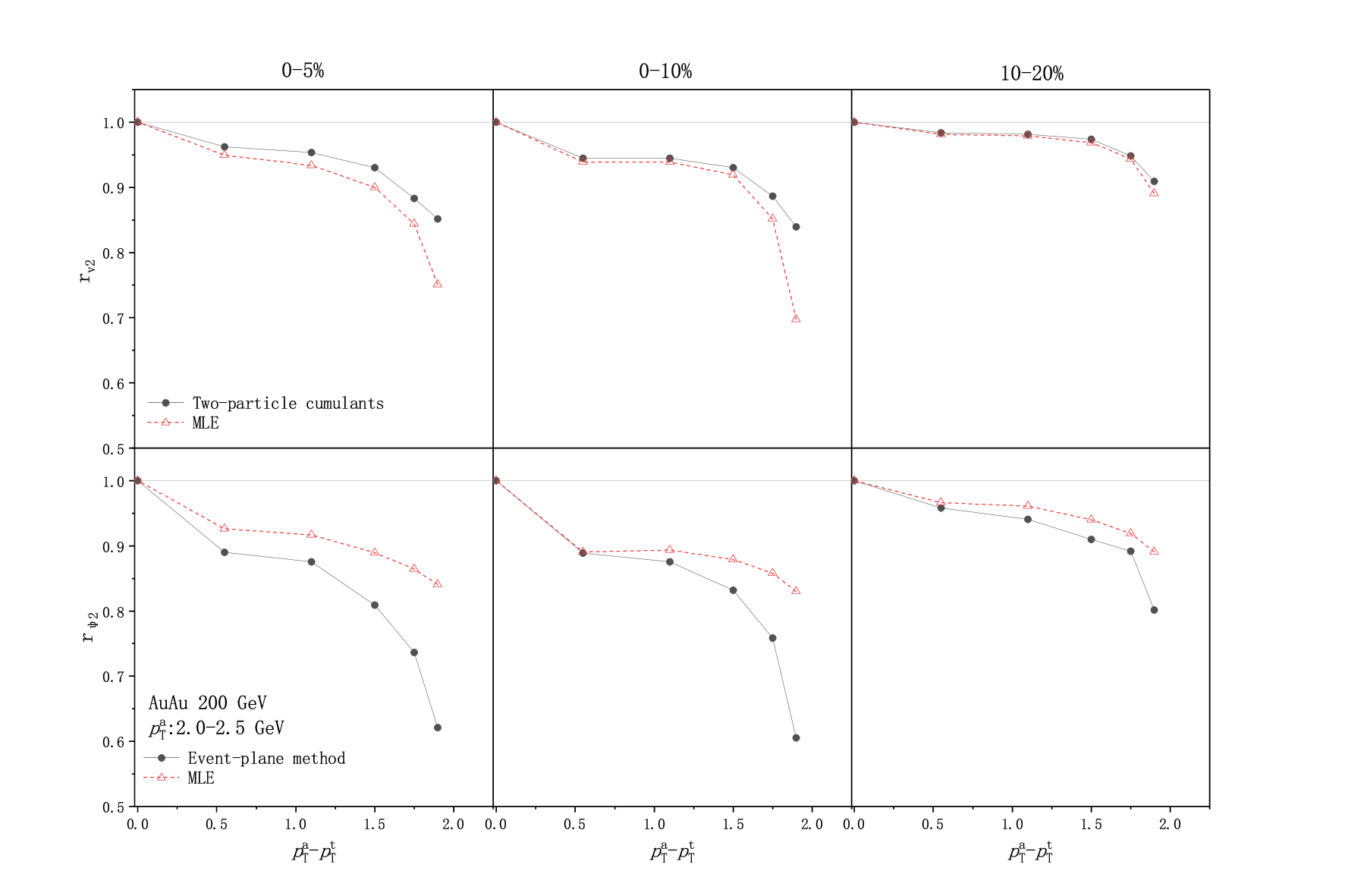}
}
\renewcommand{\figurename}{Fig.}
\renewcommand{\figurename}{Fig.}
\caption{The same as Fig.~\ref{MLECORR}, while the upper row giving a direct comparison between the multi-particle cumulants and MLE approaches.
The lower row shows the comparison between the event plane method and MLE approaches.
}
\label{vnpsin}
\end{figure}

Using the final state particles from numerical simulations with the hydrodynamic code NeXSPheRIO, we evaluate the factorization ratio $r_n$ employing both multi-particle cumulants and MLE methods.
The calculations are conducted for various centrality windows of Au+Au collisions at $\sqrt{s_{{\rm NN}}}=$ 200 GeV.
Specifically, the results for the 0-5\%, 0-10\%, and 10-20\% centrality windows are obtained by averaging over 500, 825, and 480 simulated events, respectively.
The numerical results are shown in Figs.~\ref{fig_ecntrality} to~\ref{vnpsin}.
The calculated factorization ratio is shown as a function of the difference between the transverse momenta of the trigger and associated particles.
At the origin, the two transverse momentum intervals of $p_{\rm T}^{\rm a}$ and $p_{\rm T}^{\rm t}$ coincide, and therefore, any substantial deviation from unity comes solely from the correlation within the small given interval.
Our numerical calculations indicate that the deviation from perfect factorization mostly vanishes, consistent with the experimental data~\cite{CMS:2015xmx, CMS:2013bza}.
It is understood because at the limit when the size of the interval vanishes, the Pearson correlation falls back to that of the same variable, which is guaranteed to have a perfect correlation.

We proceed to analyze the centrality dependence of the factorization breakdown in Fig.~\ref{fig_ecntrality}.
The results of $r_2$ and $r_3$ are evaluated for three different centralities, where the entire interval $2.0< p_{\rm T}^{\rm a} < 2.5$ GeV for the associated particles is integrated out.
The error bars represent the standard errors, which are insignificant compared to the difference between the two approaches.
We first focus on the results obtained by using particle cumulants.
For all cases, the factorization breakdown is observed to increase monotonically as the difference between the trigger and associated particles $|p_{\rm T}^{\rm a}-p_{\rm T}^{\rm t}|$ increases.
By comparing the three centrality windows of 0-5\%, 0-10\%, and 10-20\%, it was observed that more central collisions have a more significant factorization breakdown compared to non-central collisions.
In particular, for 0-5\% central collisions, the relative deviation in $r_2$ at the largest transverse momentum difference, $p_{\rm T}^\mathrm{a}-p_{\rm T}^\mathrm{t}\sim 2$ GeV, is about 20\%.
While for more peripheral collisions with a centrality of 10-20\%, the maximal difference in $r_2$ is only about 10\%.
These results are consistent with the experimental data and other hydrodynamic simulations, and the breakdown of factorization is attributed to the size of geometric fluctuations in the initial state.
Using the MLE method, the factorization ratio decreases with increasing momentum difference, showing a similar trend.
However, its centrality dependence is found to be rather different from that obtained by particle cumulants.
As one goes from the centrality window 0-5\% to 0-10\%, the factorization breakdown first increases, then slightly decreases or remains mainly at the same level.
The overall magnitude of the factorization evaluated using MLE is more significant than the particle cumulants method.
As discussed below, the increase in the magnitude of factorization breakdown is primarily due to more significant event-plane decorrelation, which might be partly related to the more significant event-plane fluctuations, as further elaborated below.

Next, the dependence of factorization on the specific transverse momentum intervals and different harmonic orders.
Here, we present the results for 0-5\% central collisions in Fig.~\ref{rn510} and a less central window 10-20\%, in Fig.~\ref{rn1020}.
Again, when evaluating the factorization ratio, one integrates out a specific transverse momentum interval of the associated particles.
From left to right, we consider three $p_\mathrm{T}^a$ intervals of the same width but with increasing transverse momentum.
From upper to lower, we compare two different factorization ratios, $r_2$ and $r_3$.
In general, the behavior across different centralities is found to be largely similar.
The deviation from a perfect factorization is observed to increase monotonically as the difference between the trigger and associated particles increases.
For $r_3$, the deviation becomes more significant as the transverse momentum interval of the associated particles gradually shifts to the high momentum region.
On the other hand, for $r_2$, the size of the deviation remains unchanged, largely independent of the transverse momentum intervel.
This might be due to the different physical origins of $v_2$ and $v_3$.
A similar trend is observed when compared with the results obtained using the MLE method.
The deviation from perfect correlation is more significant for the MLE results, particularly for $r_3$.
As discussed below, the deviation is mainly attributed to the more significant transverse-momentum-dependent event-plane decorrelation.


Although the above results on factorization ratios obtained by two different approaches have rather similar features, some differences are also observed.
It is meaningful to explore the origin of these deviations. 
In this regard, a possible approach is to utilize Eq.~\eqref{rn_fact_qvec}, which effectively separates the correlation originating from the flow harmonics and that associated with the event planes.
Specifically, if one assumes that event planes are largely independent of flow harmonics, we may approximate
\begin{eqnarray}
       r_n(p_\mathrm{T}^\mathrm{a}, p_\mathrm{T}^\mathrm{t}) \simeq r_{v_n}\times r_{\Psi_n} ,
\label{rn_fact_qvec2}
\end{eqnarray}
where
\begin{eqnarray}
r_{v_n}=\frac{\langle v_n(p_{\rm T}^{\rm a})v_n(p_{\rm T}^{\rm t})  \rangle}{\sqrt{\langle v_n^2(p_{\rm T}^{\rm a}) \rangle  \langle v_n^2(p_{\rm T}^{\rm t}) \rangle}}, \label{rnvn} 
\end{eqnarray}
and
\begin{eqnarray}
r_{\Psi_n}=\langle \cos n(\Psi_n(p_{\rm T}^\mathrm{a})-\Psi_n(p_{\rm T}^\mathrm{t}))\rangle , \label{rnpsin} 
\end{eqnarray}
represent the contributions from Pearson correlation of flow harmonics and event-plane correlations.

It is worth noting that the decomposition given by Eq.~\eqref{rn_fact_qvec2} is not exact, owing to the intrinsic correlation between the flow and event plane.
Nonetheless, it might still give us an educated estimation of the relative impact of individual factors.
Such an analysis is presented in Figs.~\ref{MLECORR} and~\ref{vnpsin}.
Fig.~\ref{MLECORR} shows a comparison between the quantities $r_n$, $r_{v_n}$, and $r_{\Psi_n}$ obtained using two different approaches for different centrality windows 0-5\%, 0-10\%, and 10-20\%.
The upper row of Fig.~\ref{MLECORR} represents the results calculated using multi-particle cumulants, while the lower row gives those using the MLE estimator.
It is observed that the correlations associated with flow harmonics $v_n$ and event planes $\Psi_n$ have somewhat different magnitudes while extracted using different estimators.
For the case of multi-particle cumulants, it is evident that the event plane correlation has a more pronounced impact on the resulting factorization ratio than those due to the flow harmonics.
On the other hand, the results using the MLE estimator lead to an opposite conclusion.
It is indicated that contributions from the Pearson correlation between flow harmonics and those from the event planes are essentially of the same order, while event-plane correlation is slightly more prominent.
Besides, the approximated relation Eq.~\eqref{rn_fact_qvec2} holds better regarding the magnitude for the MLE approach.

In Fig.~\ref{vnpsin}, we give a more direct comparison between the two methods regarding the flow harmonic and event-plane correlations.
The first row presents the Pearson correlation between flow harmonics calculated using the two methods.
We observe that the flow harmonic correlation largely remains the same among the two approaches. 
This is in agreement with the results on differential flow shown in Fig.~\ref{fig_differentialflow}. 
In the second row, we perform a similar comparison but for the event-plane correlations.
It is seen that the difference in event-plane correlations evaluated using the event-plane method is more significant than those obtained by the MLE approach.

\section{Factorization in terms of high-order and mixed harmonics}\label{section5}

In the literature~\cite{Bozek:2021mov}, exploring factorization using higher-order correlators has also been proposed.
Factorization ratios defined in higher-order multi-particle cumulants are expected to be more sensitive to the uncertainties in the initial conditions.
One advantage of exploring the higher-order correlators is the possibility of perfectly canceling out the event planes.
We are interested in the mixed-harmonic correlators because, by construction, event-plane cancelation is not a pertinent requirement in the context of MLE.
Therefore, one might focus exclusively on the flow correlations due to particular physical interests.

\begin{figure}[ht]
\begin{tabular}{cc}
\vspace{-26pt}
\begin{minipage}{250pt}
\centerline{\includegraphics[clip,trim=1cm 0.5cm 0 1cm, width=1.0\textwidth,height=0.8\textwidth]{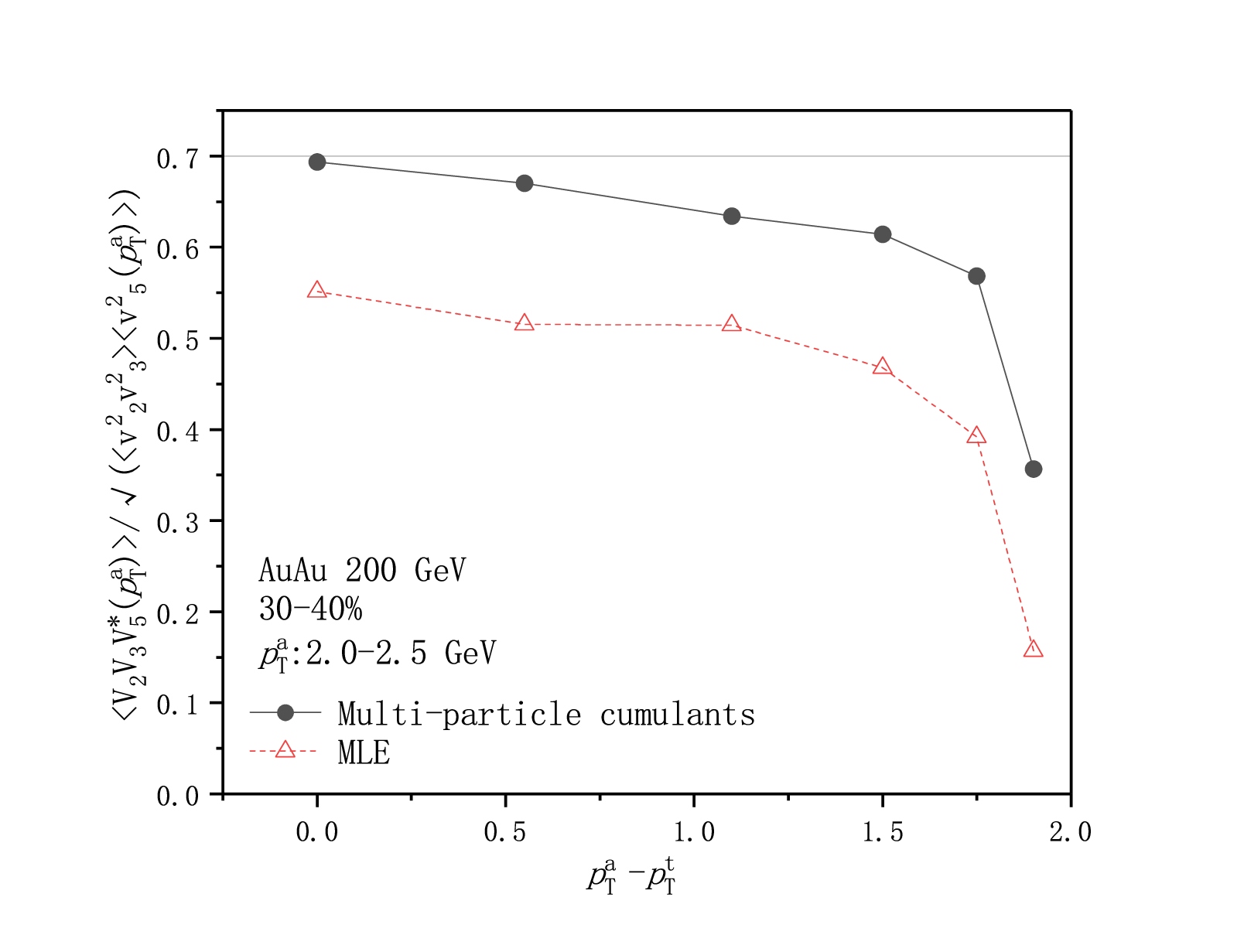}}
\end{minipage}
&
\vspace{26pt}
\begin{minipage}{250pt}
\centerline{\includegraphics[clip,trim=1cm 0.5cm 0 1cm, width=1.0\textwidth,height=0.8\textwidth]{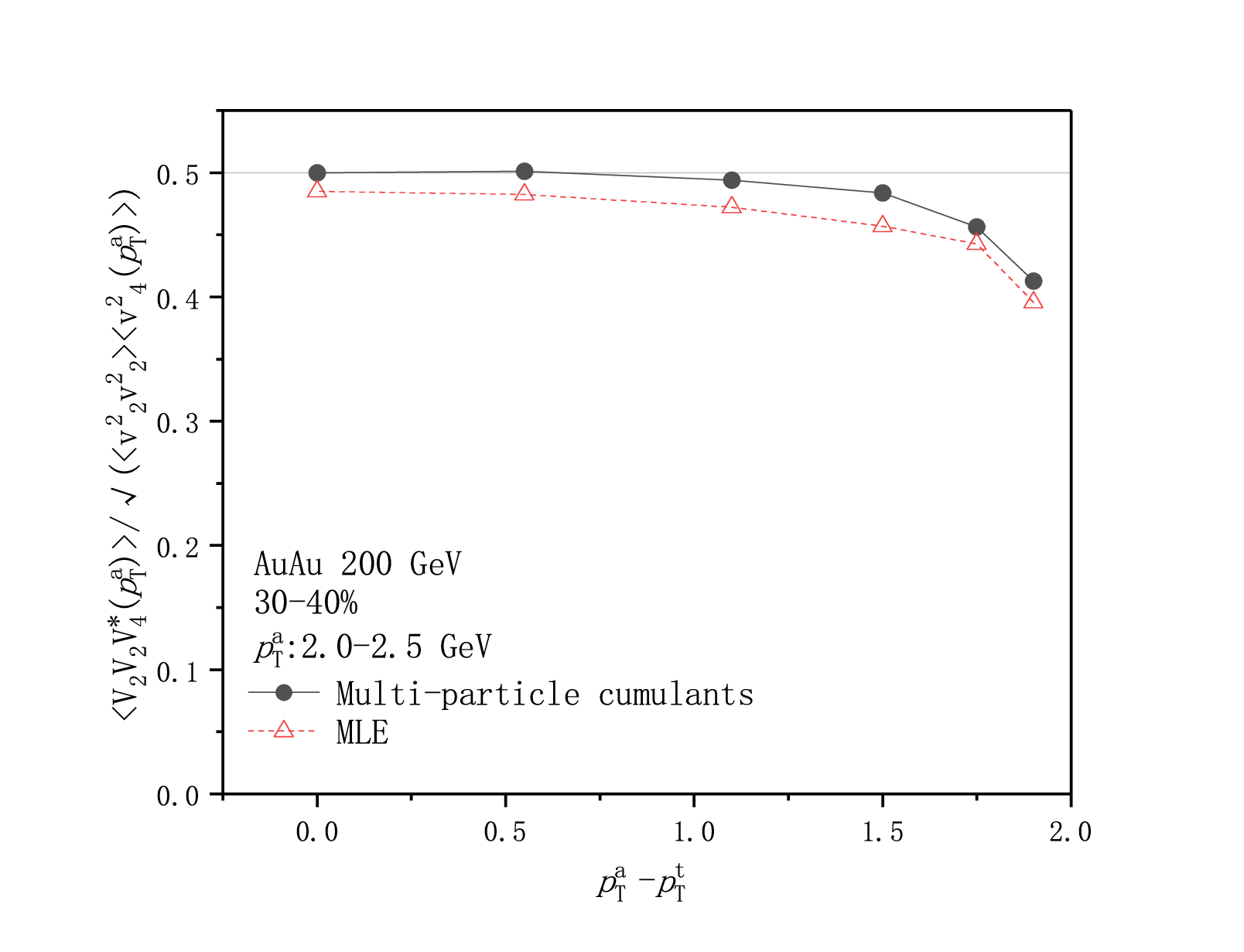}}
\end{minipage}
\end{tabular}
\renewcommand{\figurename}{Fig.}
\caption{The mixed harmonic factorization ratios evaluated using the multi-particle cumulants and MLE approaches.
The left panel shows the ratio as a function of momentum interval for $m=5$, $k=3$, and $n=2$, and the right panel gives that for $m=4$, $k=2$, and $n=2$.}
\label{vnmix}
\end{figure}

\begin{figure}[ht]
\begin{tabular}{cc}
\vspace{-26pt}
\begin{minipage}{250pt}
\centerline{\includegraphics[clip,trim=1cm 0.5cm 0 1cm, width=1.0\textwidth,height=0.8\textwidth]{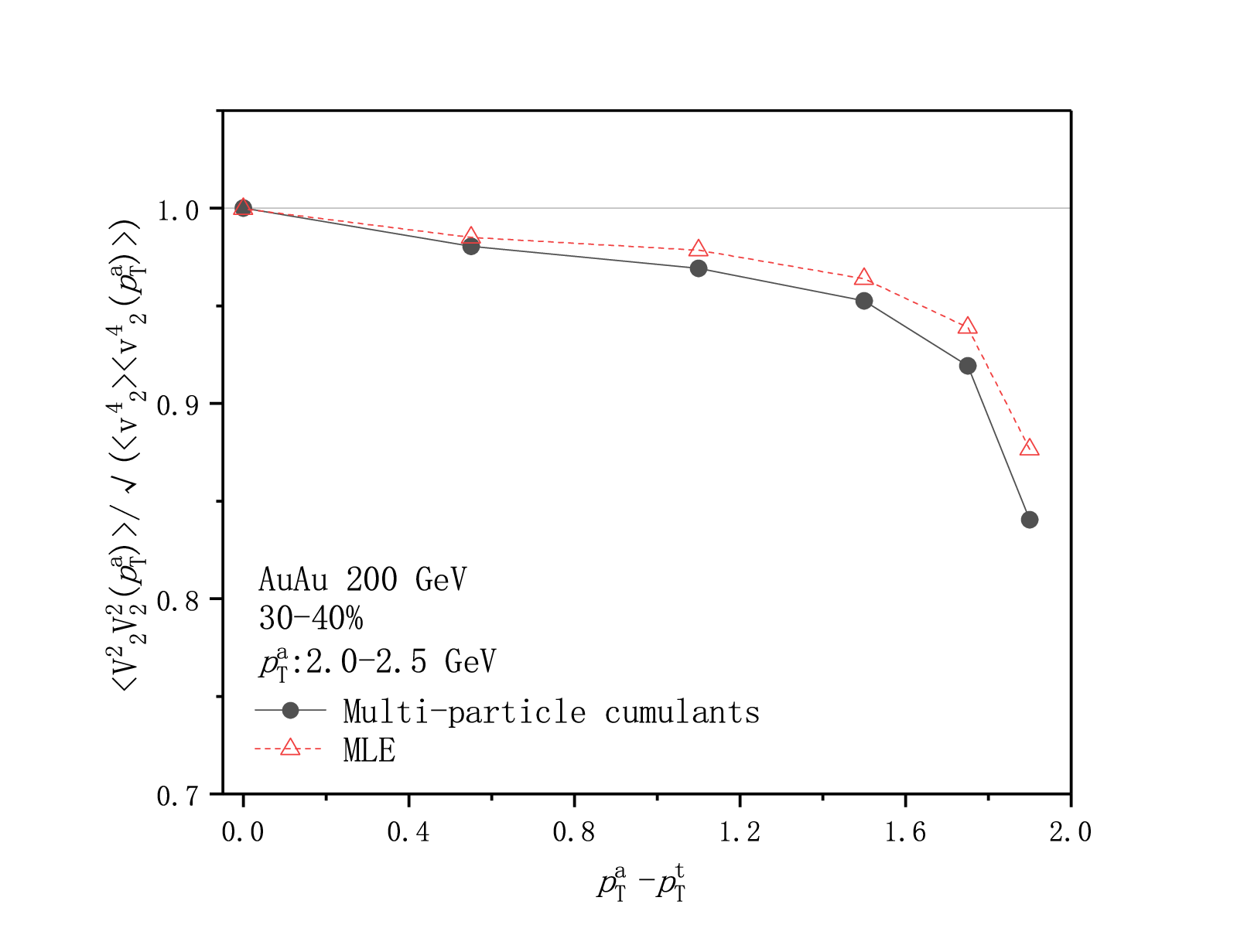}}
\end{minipage}
&
\vspace{26pt}
\begin{minipage}{250pt}
\centerline{\includegraphics[clip,trim=1cm 0.5cm 0 1cm, width=1.0\textwidth,height=0.8\textwidth]{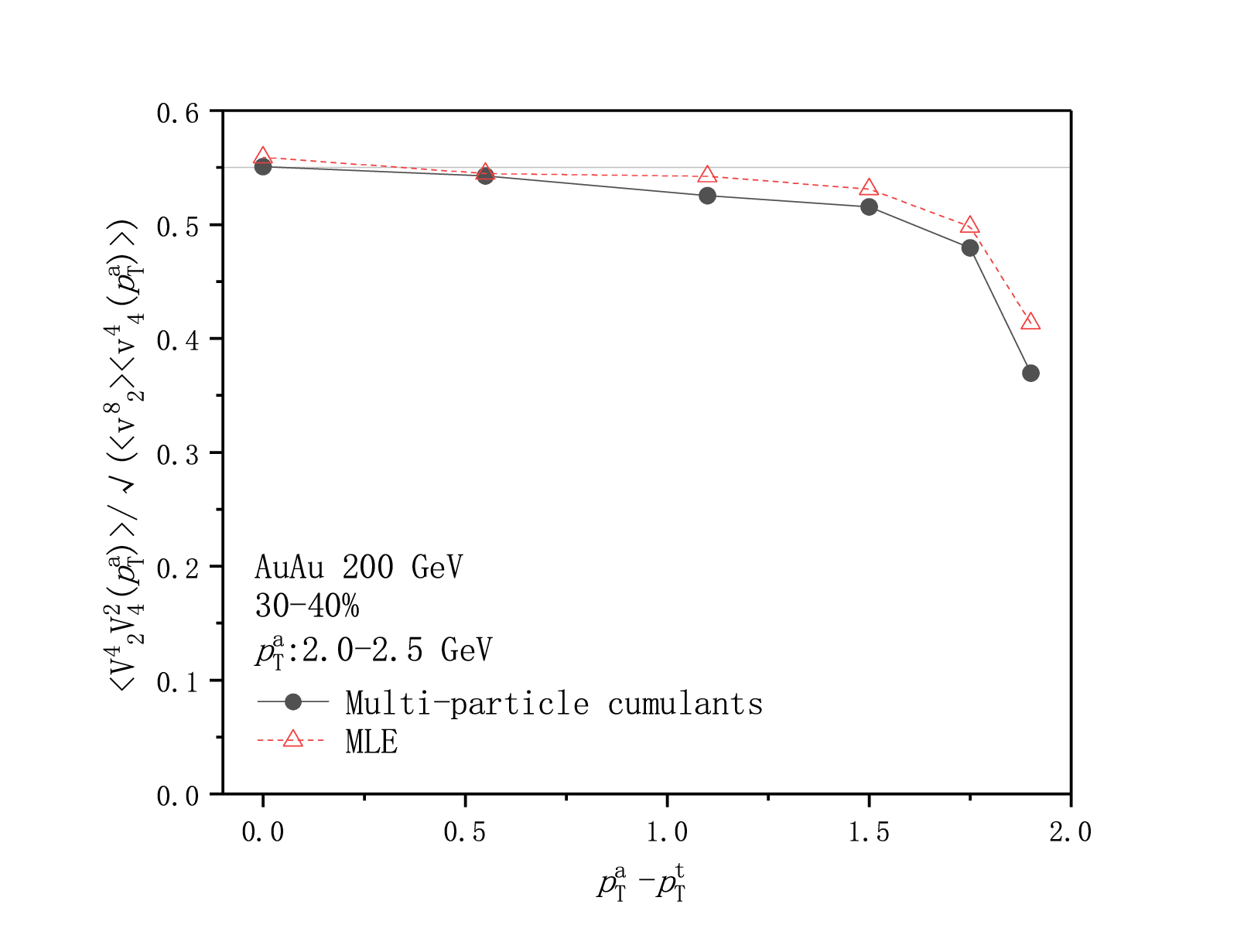}}
\end{minipage}
\end{tabular}
\renewcommand{\figurename}{Fig.}
\caption{The flow magnitude squared factorization evaluated using the multi-particle cumulants and MLE approaches.
The left panel shows the ratio as a function of momentum interval for $V_2(p_\mathrm{T}^a)$ and $V_2(p_\mathrm{T}^t)$, and the right panel gives that for $V_2^{2}(p_\mathrm{T}^a)$, and $V_4(p_\mathrm{T}^t)$.}
\label{vnmix_noEV}
\end{figure}

In this section, we will study the factorization related to the multi-particle cumulants and its further generalization to mixed harmonics of the following form
\begin{eqnarray}
r_\mathrm{mix}(p_\mathrm{T}^a, p_\mathrm{T}^t, p_\mathrm{T}^t)&=&\frac{\langle V_m^*(p_\mathrm{T}^a)V_k(p_\mathrm{T}^t)V_n(p_\mathrm{T}^t)\rangle}{\sqrt{\langle V_m^*(p_\mathrm{T}^a) V_m(p_\mathrm{T}^a)\rangle \langle V_k^*(p_\mathrm{T}^t)V_k(p_\mathrm{T}^t) V_n^*(p_\mathrm{T}^t)V_n(p_\mathrm{T}^t)\rangle}} \nb\\
&=&\frac{\langle v_m(p_\mathrm{T}^a)v_k(p_\mathrm{T}^t)v_n(p_\mathrm{T}^t) \cos{(m \Psi_m(p_\mathrm{T}^a)-k \Psi_k(p_\mathrm{T}^t)-n \Psi_n(p_\mathrm{T}^t))} \rangle}{\sqrt{\langle v_m^2(p_\mathrm{T}^a)\rangle \langle v_k^2(p_\mathrm{T}^t) v_n^2(p_\mathrm{T}^t)\rangle}},
\label{rmix} 
\end{eqnarray}
where $m=k+n$. 
Such a quantity has been investigated in the literature regarding flow fluctuations~\cite{Qian:2017ier, Bozek:2017thv}, as the term related to event-plane correlation vanishes in the case of a global constant event plane. 
It is noted that the above quantity depends on three transverse momenta. 
In practice, one integrates the associated particle's transverse momentum over a given interval, as in the previous calculations.
We are also interested in one of its variations, which is essentially the factorization where the flow magnitudes are squared, constructed in the following way~\cite{ALICE:2022dtx}
\begin{eqnarray}
r_n^{v_n^2}(p_\mathrm{T}^\mathrm{a}, p_\mathrm{T}^\mathrm{t})
=\frac{\langle V_n^*(p_\mathrm{T}^\mathrm{a})V_n(p_\mathrm{T}^\mathrm{a}) V_n^*(p_\mathrm{T}^\mathrm{t})V_n(p_\mathrm{T}^\mathrm{t})\rangle}{\sqrt{\langle V_n^{*2}(p_\mathrm{T}^\mathrm{a})V_n^{2}(p_\mathrm{T}^\mathrm{a})\rangle \langle V_n^{*2}(p_\mathrm{T}^\mathrm{t})V_n^{2}(p_\mathrm{T}^\mathrm{t})\rangle}}
=\frac{\langle v_n^2(p_\mathrm{T}^\mathrm{a}) v_n^2(p_\mathrm{T}^\mathrm{t})\rangle}{\sqrt{\langle v_n^4(p_\mathrm{T}^\mathrm{a})\rangle \langle v_n^4(p_\mathrm{T}^\mathrm{t})\rangle}}. \label{rnSquared} 
\end{eqnarray}
As indicated by the last equality of Eq.~\eqref{rnSquared}, such a quantity has the advantage of entirely canceling out the event planes by always pairing particles at some given momenta. 
In Figs.~\ref{vnmix} and~\ref{vnmix_noEV}, we explore the mixed harmonic and flow magnitude squared factorizations evaluated using the two approaches.
The analysis is carried out for the particle spectrum of 1000 events, obtained using NeXSPheRIO for the centrality window 30-40\%, for Au+Au collisions at 200 GeV.
The choice of peripheral collisions is to avoid ``noisy'' results obtained for more central collisions.

In Fig.~\ref{vnmix}, we consider two cases of mixed harmonic factorization: $m=4, k=n=2$ and $m=5, k=3, n=2$.
For such correlators, as the contribution from the event planes might not precisely cancel out, the observed difference between the two approaches is primarily attributed to the event-plane decorrelations.
It is observed that the results in both panels have a similar trend.
The factorization ratios for mixed harmonic decrease as the momentum difference increases.
The breakdown in factorization is found to be more significant for the results obtained using the MLE estimator, similar to those presented in the previous section.
In particular, the event-plane correlation does not vanish at the limit when the momenta of the trigger and associated particles coincide, giving rise to a sizable deviation from perfect factorization at the origin $p_\mathrm{T}^\mathrm{a}-p_\mathrm{T}^\mathrm{t}=0$. 
This is simply because the Pearson correlations between harmonics of different orders are not necessarily perfect.
Meanwhile, the deviation between the two approaches is less pronounced, as shown in the right panel of Fig.~\ref{vnmix}.
This is understood as $\Psi_2$ and $\Psi_4$ are largely correlated, and subsequently, the related event plane correlation is largely suppressed.

As shown in Fig.~\ref{vnmix_noEV}, similar trends are observed across different approaches for the magnitude squared factorizations.
The difference between the two approaches is less significant than what was observed in Figs.~\ref{fig_ecntrality},~\ref{rn510}, and~\ref{rn1020}.
This is primarily because, for such a scenario, the event-plane correlation does not play a role.
Subsequently, the observed agreement can be interpreted as a further indication that the Pearson correlation between flow harmonics is mainly consistent when derived using the two estimators.
The more significant deviations observed for the remaining factorization ratios are attributed primarily to the event-plane decorrelation, which is not entirely canceled out in a general case.

Up to this point, by construction, the specific forms of the correlators discussed above are dictated mainly by the cancelation of event planes.
Specifically, Eq.~\eqref{sumRes} implies a restriction on feasible types of correlators, which is irrelevant in the case of the MLE approach. 
Before closing this section, we explore a few mixed harmonic factorizations not constrained by Eq.~\eqref{sumRes}.
Specifically, we consider mixed harmonic factorization of the form,
\begin{eqnarray}
r_{v_n}(p_\mathrm{T}^a, p_\mathrm{T}^t)=\frac{\langle v_m(p_\mathrm{T}^a)v_n(p_\mathrm{T}^t)\rangle}{\sqrt{\langle v_m^2(p_\mathrm{T}^a)\rangle \langle  v_n^2(p_\mathrm{T}^t)\rangle}},
\label{rmixMLE2} 
\end{eqnarray}
where $m\neq n$, and
\begin{eqnarray}
r_{v_n}(p_\mathrm{T}^a, p_\mathrm{T}^t, p_\mathrm{T}^t)=\frac{\langle v_m(p_\mathrm{T}^a)v_k(p_\mathrm{T}^t)v_n(p_\mathrm{T}^t)\rangle}{\sqrt{\langle v_m^2(p_\mathrm{T}^a)\rangle \langle v_k^2(p_\mathrm{T}^t) v_n^2(p_\mathrm{T}^t)\rangle}},
\label{rmixMLE3} 
\end{eqnarray}
where $m\ne k+n$.
On the one hand, such quantities can be readily evaluated using MLE. 
On the other hand, inspired by Eq.~\eqref{rnSquared}, one can also evaluate similar quantities using high-order correlators where the event planes are entirely canceled out.
However, the latter factorization ratios reflect high-order moments of flow harmonics, which might be quite different from their low-order counterparts.

\begin{figure}[ht]
\begin{tabular}{cc}
\vspace{-26pt}
\begin{minipage}{250pt}
\centerline{\includegraphics[clip,trim=1cm 0.5cm 0 1cm, width=1.0\textwidth,height=0.8\textwidth]{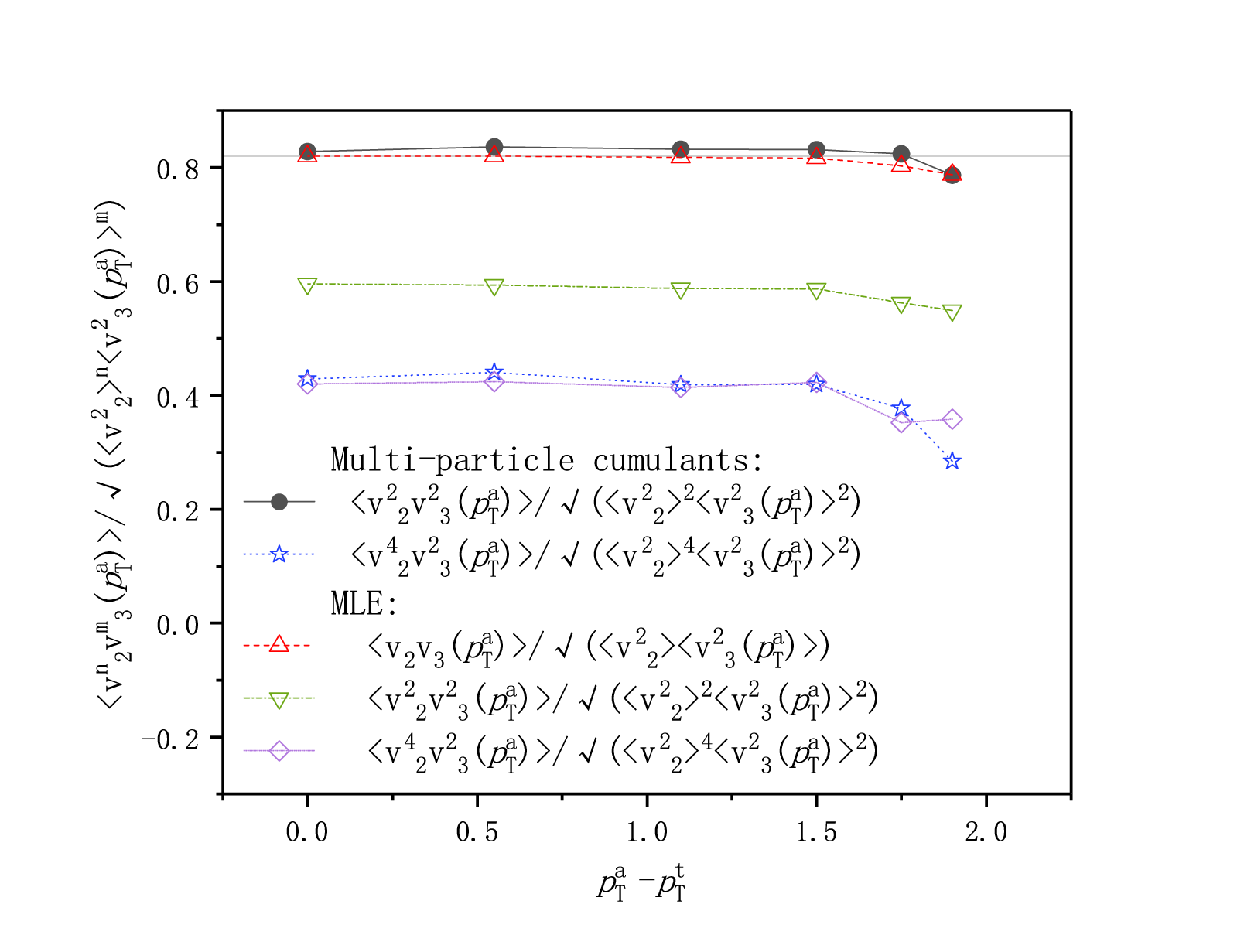}}
\end{minipage}
&
\vspace{26pt}
\begin{minipage}{250pt}
\centerline{\includegraphics[clip,trim=1cm 0.5cm 0 1cm, width=1.0\textwidth,height=0.8\textwidth]{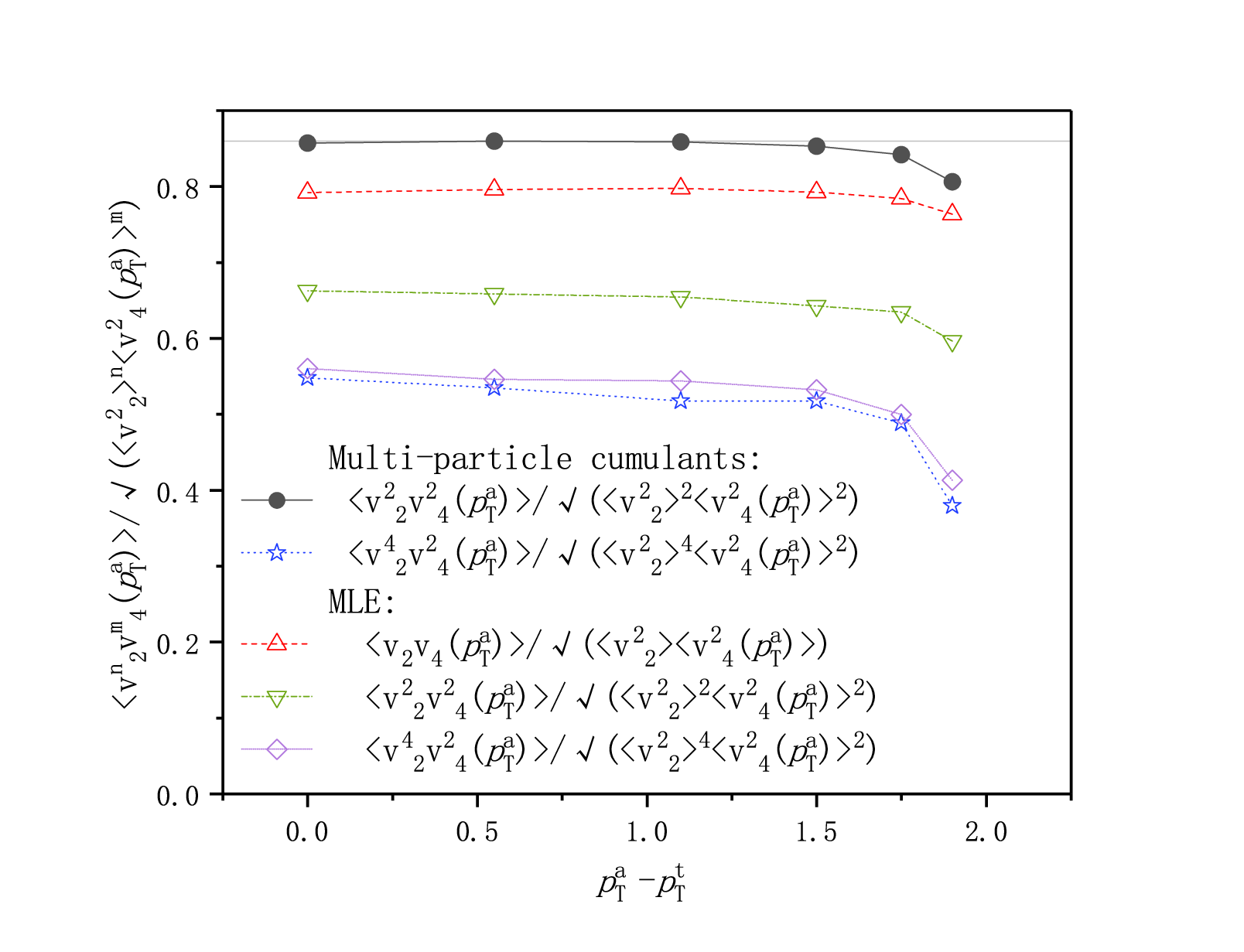}}
\end{minipage}
\\
\begin{minipage}{250pt}
\centerline{\includegraphics[clip,trim=1cm 0.5cm 0 1cm, width=1.0\textwidth,height=0.8\textwidth]{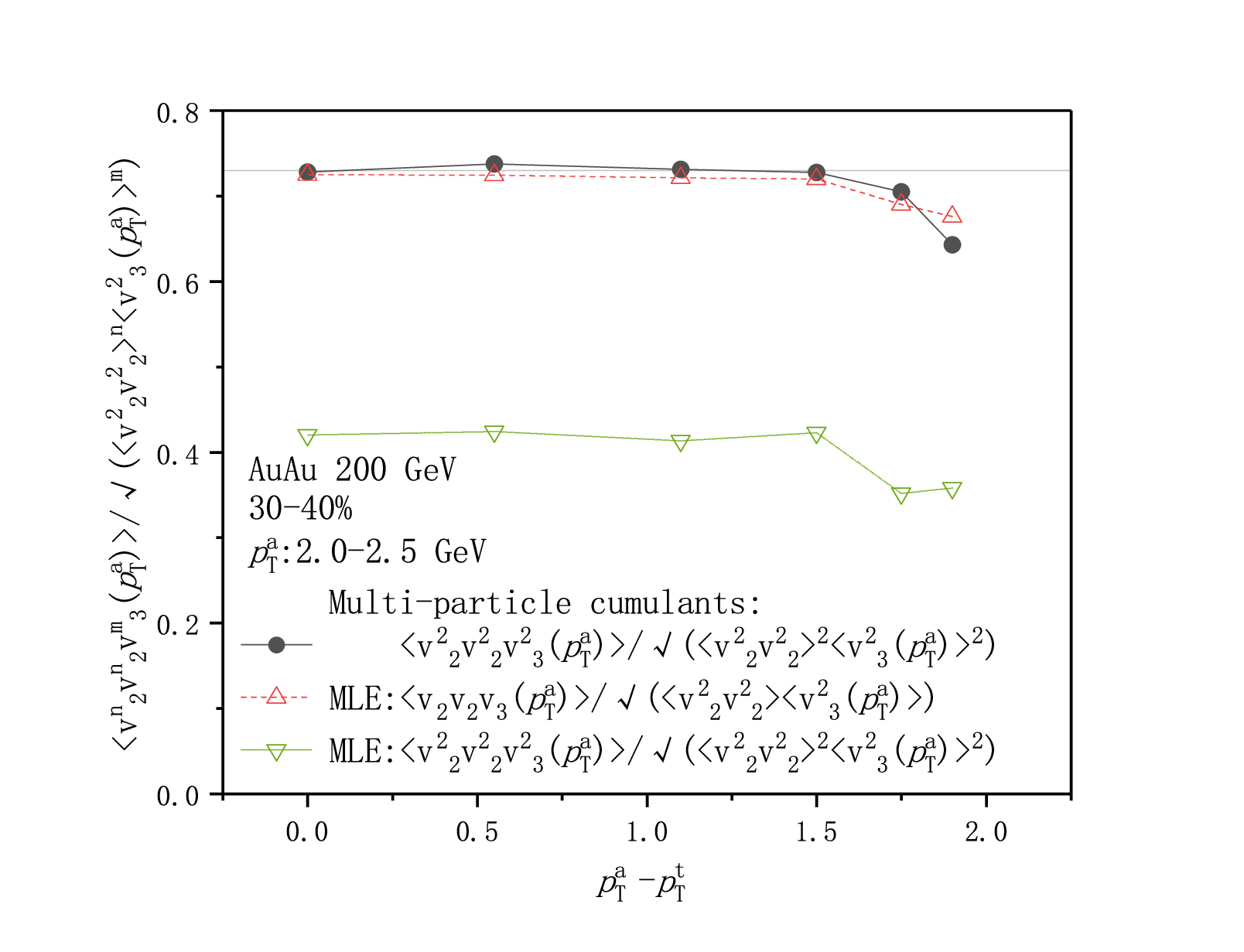}}
\end{minipage}
&
\vspace{26pt}
\begin{minipage}{250pt}
\centerline{\includegraphics[clip,trim=1cm 0.5cm 0 1cm, width=1.0\textwidth,height=0.8\textwidth]{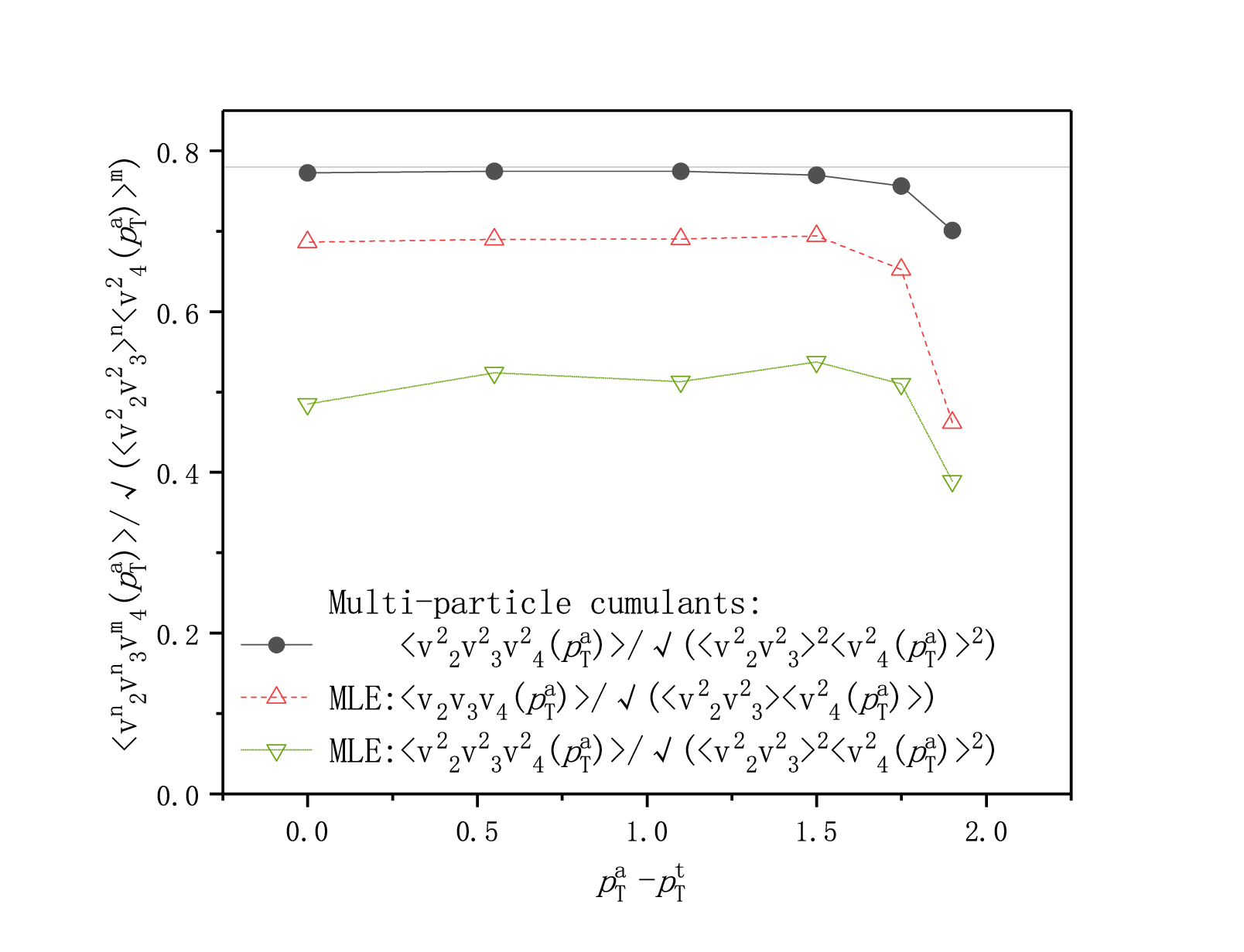}}
\end{minipage}
\end{tabular}
\renewcommand{\figurename}{Fig.}
\caption{Two- and three-particle flow harmonic correlation ratios are explored in cases where the coefficients preceding the azimuthal angles do not satisfy the condition given by Eq.~\eqref{sumRes}.
The black solid circles and blue open stars indicate the results obtained through multi-particle cumulants, while the red open triangles, green inverted open triangles, and purple open diamonds represent the outcomes derived from the MLE method.
In the upper row, the black solid circles correspond to results from four-particle cumulants, whereas the blue open stars denote results from six-particle cumulants.
In the lower row, the black solid circles depict results obtained using six-particle cumulants.}
\label{vnmix_pMLE}
\end{figure}

The numerical results are shown in Fig.~\ref{vnmix_pMLE}, where we relegate the specific forms of the differential multi-particle cumulants to Appx.~\ref{appA}.
In addition to the similar trend, a relatively sizable difference is observed between flow Pearson correlations constructed using different moments.
In other words, even though one may use high-order moments to construct appropriately normalized factorization ratios as estimators for the quantities such as Eqs.~\eqref{rmixMLE2} and~\eqref{rmixMLE3}, one observes that the deviation might be substantial, as demonstrated, for instance, by the black solid  and blue dotted curves shown in the first row of Fig.~\ref{vnmix_pMLE}.
Specifically, the deviation from perfect factorization increases with the order.
Owing to the {\it equivariance} of MLE~\cite{book-statistical-inference-Wasserman}, such estimations can also be readily generalized for higher moments.
As a comparison, these results are also presented in Fig.~\ref{vnmix_pMLE}.
Once again, a sizable difference is observed between these two approaches, for instance, by the black solid and green dashed curves shown in the first row of Fig.~\ref{vnmix_pMLE}.
The above results indicate that the consistency between different approaches observed in differential flow and Pearson correlation of flow harmonics obtained using lowest-order moments does not generalize to the higher-order scenario.
In other words, the observables, such as factorization ratios, furnished by higher-order moments become more sensitive to the underlying statistical estimators.
In this regard, MLE provides an intuitive alternative to assess collective flows than the more commonly employed methods.

\section{Concluding remarks}\label{section6}

This work explores several aspects of applying MLE to differential flow, flow factorization, and event-plane correlations.
We explored the application of MLE to calculate differential flow using both simulated and experimental data. 
The numerical results indicate that MLE gives consistent results on the differential flow compared to those obtained using the standard methods.
Our analysis proceeds to more sensitive observables, such as flow factorization.
Also, we investigate the dependence of the factorization on different centralities and harmonic orders.
Although the primary features are similar, the results indicate a substantial difference in these quantities obtained using cumulants and MLE approaches.
To better understand the underlying physics of the observed flow factorization, we explore the constituting flow harmonic and event-plane correlations.
The observed difference in event-plane correlation is understood to give rise to the resultant discrepancies in flow factorization.
We also study mixed harmonic and higher-order factorizations.
In particular, by construction, only some specific higher-order harmonic correlators are entirely independent of event-plane correlation, while the MLE is not affected by such constraints.
The higher-order flow factorization ratios are found to be more sensitive than their lower-order counterparts, potentially providing crucial information on the initial-state geometric fluctuations.
We therefore argue that the MLE estimator furnishes a relevant alternative to such analysis.

As an asymptotically normal estimator, the efficiency of MLE is well established.
A primary challenge of the approach is its computational cost.
The present study generalizes our initial proposal to broader applications and shows that such an extension is indeed computationally feasible. 
In this regard, the flexibility of the MLE, such as the fact that it does not rely on the specific construction of particle pairs or tuples to cancel event planes, can be utilized in our favor.
Compared to the standard methods, it might be more flexible to deal with flow analysis scenarios where a template for the nonflow is known to a certain degree.
We plan to continue exploring related topics in future studies.

\section*{Acknowledgements}

We are thankful for enlightened discussions with Leonardo Barbosa, Pedro Ishida, Mike Lisa, Matthew Luzum, and Giorgio Torrieri.
This work is supported by the National Natural Science Foundation of China.
We also gratefully acknowledge the financial support from Brazilian agencies 
Funda\c{c}\~ao de Amparo \`a Pesquisa do Estado de S\~ao Paulo (FAPESP), 
Fundação de Amparo à Pesquisa do Estado do Rio Grande do Sul (FAPERGS),
Funda\c{c}\~ao de Amparo \`a Pesquisa do Estado do Rio de Janeiro (FAPERJ), 
Conselho Nacional de Desenvolvimento Cient\'{\i}fico e Tecnol\'ogico (CNPq), 
and Coordena\c{c}\~ao de Aperfei\c{c}oamento de Pessoal de N\'ivel Superior (CAPES).
A part of this work was developed under the project Institutos Nacionais de Ciências e Tecnologia - Física Nuclear e Aplicações (INCT/FNA) Proc. No. 464898/2014-5.
This research is also supported by the Center for Scientific Computing (NCC/GridUNESP) of S\~ao Paulo State University (UNESP).
This work is also supported by the Postgraduate Research \& Practice Innovation Program of Jiangsu Province under Grant No. KYCX22-3453.

\appendix

\section{Explicit forms for the differential multi-particle cumulants}\label{appA}

This appendix presents the expressions for the differential multi-particle cumulants utilized in Sec.~\ref{section5}.
Although similar expressions have been derived in the literature~\cite{Bilandzic:2013kga, Pruneau:2023cea, Nielsen:2023znu, Kubo1962GENERALIZEDCE}, some particular cases utilized in Fig.~\ref{vnmix_pMLE} has to be specifically derived.

For our analysis, we divide the particles in each event into reference particles (RP) and particles of interest (POI).
We note that these two types of particles may overlap in practice.
For a specific event, RP corresponds to particles within a narrow differential bin of the trigger particles, while POI corresponds to a statistical sample of associated particles within specific $p_\mathrm{T}$ intervals.
In the differential multi-particle cumulants, we use underscores to denote particles from POIs, while the remaining particles come from RPs.

The $m$-particle correlator defined in Sec.~\ref{section2} can be generalized to the following momentum-dependent form~\cite{Bilandzic:2013kga}:
\begin{eqnarray}
     \langle m' \rangle_{\underline{n_{1}},\underline{n_{2}},n_3,\cdots ,n_{m}} 
     \equiv \langle e^{i(n_{1}\phi_{k_1}+n_{2}\phi_{k_2}+n_{3}\phi_{k_3}+\cdots+n_{m}\phi_{k_m})} \rangle ,
\label{appMultim}
\end{eqnarray}
which is intuitively given by
\begin{eqnarray}
\frac{\sum\limits_{k_1,k_2}^{M_{P}}\sum\limits_{\substack{k_3,\cdots,k_m\\k_1\ne k_2\ne k_3 \ne \cdots \ne k_m}}^{M_{R}} w_{k_1}w_{k_2}w_{k_3}\cdots w_{k_m} e^{i(n_{1}\phi_{k_1}+n_{2}\phi_{k_2}+n_{3}\phi_{k_3}+\cdots+n_{m}\phi_{k_m})}}{\sum\limits_{k_1,k_2}^{M_{P}}\sum\limits_{\substack{k_3,\cdots,k_m\\k_1\ne k_2\ne k_3 \ne \cdots \ne k_m}}^{M_{R}} w_{k_1}w_{k_2}w_{k_3}\cdots w_{k_m}}
    \equiv \frac{\mathrm{N}_{\langle m' \rangle_{\underline{n_{1}},\underline{n_{2}},n_3,\cdots ,n_{m}}}}{\mathrm{D}_{\langle m' \rangle_{\underline{n_{1}},\underline{n_{2}},n_3,\cdots ,n_{m}}}} ,
\label{appMultim_ND}
\end{eqnarray}
where $M_R$ represents the total number of particles labeled as RP in an event, and $M_P$ corresponds to the POI ones.
Additionally, $w$ gives the weight of the particles.
We use $Q$-, $p$-, and $q$-vectors to represent the vector expressions of the particles from different intervals.
The weighted $Q$-vectors is a complex number defined by
\begin{eqnarray}
    Q_{n,l}\equiv \sum_{k=1}^{M_R}w_{k}^{l}e^{in\phi_{k}} ,
\label{appQRP}
\end{eqnarray}
where the summation enumerates all the particles labeled as RPs in an event.
Similarly, the weighted $p$-vectors are constructed out of all POI-labeled particles within a momentum bin:
\begin{eqnarray}
    p_{n,l}\equiv \sum_{k=1}^{M_P}w_{k}^{l}e^{in\phi_{k}} .
\label{appQPOI}
\end{eqnarray}
Finally, the weighted q-vectors consist only of particles from the differential bin of RPs in the event, which are particles marked as both RP and POI:
\begin{eqnarray}
    q_{n,l}\equiv \sum_{k=1}^{M_q}w_{k}^{l}e^{in\phi_{k}} ,
\label{appQq}
\end{eqnarray}
where the total number of such particles is $M_q$.

Given the above definitions, one finds the following expressions for differential 2-, 3-, and 4-particle correlations:
\begin{eqnarray}
     \mathrm{N}_{\langle 2\rangle _{\underline{n_1}, n_2}} &=& p_{n_1,1}Q_{n_2,1}-q_{n_1+n_2,2} , \label{app2pN}\\
     \mathrm{D}_{\langle 2\rangle _{\underline{n_1}, n_2}} &=& p_{0,1}Q_{0,1}-q_{0,2} \label{app2pD} .
\label{app2p}
\end{eqnarray}
\begin{eqnarray}
     \mathrm{N}_{\langle 3\rangle _{\underline{n_1}, n_2, n_3}} &=& p_{n_1,1}Q_{n_2,1}Q_{n_3,1}-q_{n_1+n_2,2}Q_{n_3,1}-q_{n_1+n_3,2}Q_{n_2,1} \nonumber\\
     & &-p_{n_1,1}Q_{n_2+n_3,2}+2q_{n_1+n_2+n_3,3}  , \label{app3pN}\\
     \mathrm{D}_{\langle 3\rangle _{\underline{n_1}, n_2, n_3}} &=& p_{0,1}Q_{0,1}^2-p_{0,1}Q_{0,2}-2q_{0,2}Q_{0,1}+2q_{0,3} \label{app3pD} .
\label{app3p}
\end{eqnarray}
\begin{eqnarray}
     \mathrm{N}_{\langle 4\rangle _{\underline{n_1}, \underline{n_2}, n_3, n_4}} &=&  p_{n_1,1}p_{n_2,1}Q_{n_3,1}Q_{n_4,1}-p_{n_1+n_2,2}Q_{n_3,1}Q_{n_4,1}-q_{n_1+n_3,2}p_{n_2,1}Q_{n_4,1}\nonumber\\
     & & -p_{n_1,1}q_{n_2+n_3,2}Q_{n_4,1}+2q_{n_1+n_2+n_3,3}Q_{n_4,1}-q_{n_1+n_4,2}p_{n_2,1}Q_{n_3,1}\nonumber\\
     & & +q_{n_1+n_4,2}q_{n_2+n_3,2}-p_{n_1,1}Q_{n_3,1}q_{n_2+n_4,2}+q_{n_1+n_3,2}q_{n_2+n_4,2}\nonumber\\
     & & +2q_{n_1+n_2+n_4,3}Q_{n_3,1}-p_{n_1,1}p_{n_2,1}Q_{n_3+n_4,2}+p_{n_1+n_2,2}Q_{n_3+n_4,2}\nonumber\\
     & & +2q_{n_1+n_3+n_4,3}p_{n_2,1}+2p_{n_1,1}q_{n_2+n_3+n_4,3}-6q_{n_1+n_2+n_3+n_4,4}
     , \label{app4pN}\\
     \mathrm{D}_{\langle 4\rangle _{\underline{n_1}, \underline{n_2}, n_3, n_4}} &=& p_{0,1}^2Q_{0,1}^2-p_{0,2}Q_{0,1}^2-4q_{0,2}p_{0,1}Q_{0,1}+4q_{0,3}Q_{0,1}+2q_{0,2}^2-p_{0,1}^2Q_{0,2}\nonumber\\
     & &  +4q_{0,3}p_{0,1}-p_{0,2}Q_{0,2}+6q_{0,4} \label{app4pD} .
\label{app4p}
\end{eqnarray}
\begin{eqnarray}
     \mathrm{N}_{\langle 6\rangle _{\underline{n_1}, \underline{n_2}, n_3, n_4, n_5, n_6}} &=&  p_{n_1,1}p_{n_2,1}Q_{n_3,1}Q_{n_4,1}Q_{n_5,1}Q_{n_6,1}-q_{n_1+n_2+n_3+n_4+n_5,5}Q_{n_6,1}-q_{n_1+n_2+n_3+n_4+n_6,5}Q_{n_5,1}\nonumber\\
     & & -q_{n_1+n_2+n_3+n_5+n_6,5}Q_{n_4,1}-q_{n_1+n_2+n_4+n_5+n_6,5}Q_{n_3,1}-q_{n_1+n_3+n_4+n_5+n_6,5}p_{n_2,1}\nonumber\\
     & &  -q_{n_2+n_3+n_4+n_5+n_6,5}p_{n_1,1}-\sum_{(15)}(q_{n_1+n_2+n_3+n_4,4}Q_{n_5,1}Q_{n_6,1}+q_{n_1+n_2+n_3+n_5,4}Q_{n_4,1}Q_{n_6,1}\nonumber\\
     & & +\cdots+-q_{n_1+n_2+n_5+n_6,4}Q_{n_3,1}Q_{n_4,1})-\sum_{(15)}(q_{n_1+n_2+n_3+n_4,4}Q_{n_5+n_6,2}+q_{n_1+n_2+n_3+n_5,4}Q_{n_4+n_6,2}\nonumber\\
     & & +\cdots+q_{n_1+n_2+n_5+n_6,4}Q_{n_3+n_4,2})-\sum_{(10)}(q_{n_1+n_2+n_3,3}Q_{n_4+n_5+n_6,3}+q_{n_1+n_2+n_4,3}Q_{n_3+n_5+n_6,3}\nonumber\\ 
     & & +\cdots+q_{n_1+n_3+n_6,3}q_{n_2+n_4+n_5,3})+2\sum_{(60)}(q_{n_1+n_2+n_3,3}Q_{n_4+n_5,2}Q_{n_6,1}\nonumber\\
     & & +q_{n_1+n_2+n_3,3}Q_{n_4+n_6,2}Q_{n_5,1}+\cdots+Q_{n_4+n_5+n_6,3}p_{n_1+n_2,2}Q_{n_3,1})\nonumber\\
     & &)-6\sum_{(20)}(q_{n_1+n_2+n_3,3}Q_{n_4,1}Q_{n_5,1}Q_{n_6,1}+q_{n_1+n_2+n_4,3}Q_{n_3,1}Q_{n_5,1}Q_{n_6,1}\nonumber\\
     & & + \cdots+Q_{n_4+n_5+n_6,3}p_{n_1,1}p_{n_2,1}Q_{n_3,1})+2\sum_{(15)}(p_{n_1+n_2,2}Q_{n_3+n_4,2}Q_{n_5+n_6,2}\nonumber\\
     & & +p_{n_1+n_2,2}Q_{n_3+n_5,2}Q_{n_4+n_6,2}+ \cdots+Q_{n_5+n_6,2}q_{n_1+n_3,2}q_{n_2+n_4,2})\nonumber\\
     & &-6\sum_{(45)}(p_{n_1+n_2,2}Q_{n_3+n_4,2}Q_{n_5,1}Q_{n_6,1}+p_{n_1+n_2,2}Q_{n_3+n_5,2}Q_{n_4,1}Q_{n_6,1}\nonumber\\
     & &+\cdots+Q_{n_5+n_6,2}q_{n_1+n_3,2}p_{n_2,1}Q_{n_4,1})+24\sum_{(15)}(p_{n_1+n_2,2}Q_{n_3,1}Q_{n_4,1}Q_{n_5,1}Q_{n_6,1}\nonumber\\
     & &+q_{n_1+n_3,2}p_{n_2,1}Q_{n_4,1}Q_{n_5,1}Q_{n_6,1}+\cdots+Q_{n_5+n_6,2}p_{n_1,1}p_{n_2,1}Q_{n_3,1}Q_{n_4,1})\nonumber\\
     & &-120(q_{n_1+n_2+n_3+n_4+n_5+n_6}),\\
     \mathrm{D}_{\langle 6\rangle _{\underline{n_1}, \underline{n_2}, n_3, n_4, n_5, n_6}}
     &=& p_{0,1}^2Q_{0,1}^4-6q_{0,4}Q_{0,1}-6q_{0,4}Q_{0,1}^2-8q_{0,4}Q_{0,1}^2-Q_{0,4}p_{0,1}^2-Q_{0,4}p_{0,1}^2-5q_{0,4}Q_{0,2}\nonumber\\
     & &  -9q_{0,4}q_{0,2}-Q_{0,4}q_{0,2}-7q_{0,3}Q_{0,3}-3q_{0,3}q_{0,3}+2(12q_{0,3}Q_{0,2}Q_{0,1}+24q_{0,3}q_{0,2}Q_{0,1}\nonumber\\
     & & +12q_{0,3}Q_{0,2}p_{0,1}+4Q_{0,3}p_{0,2}Q_{0,1}+8Q_{0,3}q_{0,2}p_{0,1})-6(4q_{0,3}Q_{0,1}^3+12q_{0,3}p_{0,1}Q_{0,1}^2\nonumber\\
     & &  +4q_{0,3}p_{0,1}^2Q_{0,1})+2(3p_{0,2}Q_{0,2}^2+12q_{0,2}^2Q_{0,2})-6(3p_{0,2}Q_{0,2}Q_{0,1}^2+24q_{0,2}^2Q_{0,1}^2\nonumber\\
     & & +12Q_{0,2}{q_{0,2}p_{0,1}Q_{0,1}+6Q_{0,2}p_{0,2}Q_{0,1}^2})\nonumber\\
     & &  +24(p_{0,2}Q_{0,1}^4+8q_{0,2}p_{0,1}Q_{0,1}^3+6Q_{0,2}p_{0,1}^2Q_{0,1}^2)-120q_{0,6} \label{app6pN} .
\label{app6pD}
\end{eqnarray}
In deriving the six-particle correlator, one has performed several summations to reduce some of the repetitive terms.
The subscript in the summation symbol represents the total number of terms, and all the remaining terms in the summation are obtained by rotating through different particle indices $n_1$ to $n_6$.
The above formulas can be readily applied to all possible scenarios, inclusively when RP and POI particles do not overlap or fully overlap.

\bibliographystyle{h-physrev}
\bibliography{references_MLE, references_qian}

\end{document}